\documentclass[12pt]{article}
\usepackage{graphicx}
\usepackage{lscape}
\usepackage{citesort}
\usepackage{amssymb}
\usepackage{appendix}
\usepackage{multirow}

\begin{document}
\def\qq{\langle \bar q q \rangle}
\def\uu{\langle \bar u u \rangle}
\def\dd{\langle \bar d d \rangle}
\def\sp{\langle \bar s s \rangle}
\def\GG{\langle g_s^2 G^2 \rangle}
\def\Tr{\mbox{Tr}}
\def\figt#1#2#3{
        \begin{figure}
        $\left. \right.$
        \vspace*{-2cm}
        \begin{center}
        \includegraphics[width=10cm]{#1}
        \end{center}
        \vspace*{-0.2cm}
        \caption{#3}
        \label{#2}
        \end{figure}
	}
	
\def\figb#1#2#3{
        \begin{figure}
        $\left. \right.$
        \vspace*{-1cm}
        \begin{center}
        \includegraphics[width=10cm]{#1}
        \end{center}
        \vspace*{-0.2cm}
        \caption{#3}
        \label{#2}
        \end{figure}
                }

\def\ds{\displaystyle}
\def\beq{\begin{equation}}
\def\eeq{\end{equation}}
\def\bea{\begin{eqnarray}}
\def\eea{\end{eqnarray}}
\def\beeq{\begin{eqnarray}}
\def\eeeq{\end{eqnarray}}
\def\ve{\vert}
\def\vel{\left|}
\def\ver{\right|}
\def\nnb{\nonumber}
\def\ga{\left(}
\def\dr{\right)}
\def\aga{\left\{}
\def\adr{\right\}}
\def\lla{\left<}
\def\rra{\right>}
\def\rar{\rightarrow}
\def\lrar{\leftrightarrow}  
\def\nnb{\nonumber}
\def\la{\langle}
\def\ra{\rangle}
\def\ba{\begin{array}}
\def\ea{\end{array}}
\def\tr{\mbox{Tr}}
\def\ssp{{\Sigma^{*+}}}
\def\sso{{\Sigma^{*0}}}
\def\ssm{{\Sigma^{*-}}}
\def\xis0{{\Xi^{*0}}}
\def\xism{{\Xi^{*-}}}
\def\qs{\la \bar s s \ra}
\def\qu{\la \bar u u \ra}
\def\qd{\la \bar d d \ra}
\def\qq{\la \bar q q \ra}
\def\gGgG{\la g^2 G^2 \ra}
\def\q{\gamma_5 \not\!q}
\def\x{\gamma_5 \not\!x}
\def\g5{\gamma_5}
\def\sb{S_Q^{cf}}
\def\sd{S_d^{be}}
\def\su{S_u^{ad}}
\def\sbp{{S}_Q^{'cf}}
\def\sdp{{S}_d^{'be}}
\def\sup{{S}_u^{'ad}}
\def\ssp{{S}_s^{'??}}

\def\sig{\sigma_{\mu \nu} \gamma_5 p^\mu q^\nu}
\def\fo{f_0(\frac{s_0}{M^2})}
\def\ffi{f_1(\frac{s_0}{M^2})}
\def\fii{f_2(\frac{s_0}{M^2})}
\def\O{{\cal O}}
\def\sl{{\Sigma^0 \Lambda}}
\def\es{\!\!\! &=& \!\!\!}
\def\ap{\!\!\! &\approx& \!\!\!}
\def\ar{&+& \!\!\!}
\def\ek{&-& \!\!\!}
\def\kek{\!\!\!&-& \!\!\!}
\def\cp{&\times& \!\!\!}
\def\se{\!\!\! &\simeq& \!\!\!}
\def\eqv{&\equiv& \!\!\!}
\def\kpm{&\pm& \!\!\!}
\def\kmp{&\mp& \!\!\!}
\def\mcdot{\!\cdot\!}
\def\erar{&\rightarrow&}

% .........................................................

\def\simlt{\stackrel{<}{{}_\sim}}
\def\simgt{\stackrel{>}{{}_\sim}}

% .........................................................

\renewcommand{\textfraction}{0.2}    %float (figures) parameters
\renewcommand{\topfraction}{0.8}   

\renewcommand{\bottomfraction}{0.4}   
\renewcommand{\floatpagefraction}{0.8}
\newcommand\mysection{\setcounter{equation}{0}\section}

\def\baeq{\begin{appeq}}     \def\eaeq{\end{appeq}}  
\def\baeeq{\begin{appeeq}}   \def\eaeeq{\end{appeeq}}
\newenvironment{appeq}{\beq}{\eeq}   
\newenvironment{appeeq}{\beeq}{\eeeq}
\def\bAPP#1#2{
 \markright{APPENDIX #1}
 \addcontentsline{toc}{section}{Appendix #1: #2}
 \medskip
 \medskip
 \begin{center}      {\bf\LARGE Appendix #1 :}{\quad\Large\bf #2}
% \begin{center}      {\bf\LARGE Appendix  :}{\quad\Large\bf #2}
\end{center}
 \renewcommand{\thesection}{#1.\arabic{section}}
\setcounter{equation}{0}
        \renewcommand{\thehran}{#1.\arabic{hran}}
\renewenvironment{appeq}
  {  \renewcommand{\theequation}{#1.\arabic{equation}}
     \beq
  }{\eeq}
\renewenvironment{appeeq}
  {  \renewcommand{\theequation}{#1.\arabic{equation}}
     \beeq
  }{\eeeq}
\nopagebreak \noindent}

\def\eAPP{\renewcommand{\thehran}{\thesection.\arabic{hran}}}

\renewcommand{\theequation}{\arabic{equation}}
\newcounter{hran}
\renewcommand{\thehran}{\thesection.\arabic{hran}}

\def\bmini{\setcounter{hran}{\value{equation}}
\refstepcounter{hran}\setcounter{equation}{0}
\renewcommand{\theequation}{\thehran\alph{equation}}\begin{eqnarray}}
\def\bminiG#1{\setcounter{hran}{\value{equation}}
\refstepcounter{hran}\setcounter{equation}{-1}
\renewcommand{\theequation}{\thehran\alph{equation}}
\refstepcounter{equation}\label{#1}\begin{eqnarray}}

%       the stuff below defines \eqalign and \eqalignno in such a
%       way that they will run on Latex

\newskip\humongous \humongous=0pt plus 1000pt minus 1000pt
\def\caja{\mathsurround=0pt}
%\def\eqalign#1{\,\vcenter{\openup1\jot
%\caja   %\ialign{\strut \hfil$\displaystyle{##}$&$
%\displaystyle{{}##}$\hfil\crcr#1\crcr}
%}\,}

% ...........................................................

\title{
         {\Large
                 {\bf
Light vector meson octet--decuplet baryon vertices in QCD
                 }
         }
      }

\author{\vspace{1cm}\\
{\small T. M. Aliev$^a$ \thanks
{e-mail: taliev@metu.edu.tr}~\footnote{permanent address:Institute
of Physics,Baku,Azerbaijan}\,\,,
A. \"{O}zpineci$^a$ \thanks
{e-mail: ozpineci@p409a.physics.metu.edu.tr}\,\,,
M. Savc{\i}$^a$ \thanks
{e-mail: savci@metu.edu.tr} \,\,,
V. S. Zamiralov$^b$ \thanks  
{e-mail: zamir@depni.sinp.msu.ru}} \\
{\small (a) Physics Department, Middle East Technical University,
06531 Ankara, Turkey} \\
{\small (b) Institute of Nuclear Physics, M. V. Lomonosov MSU, Moscow,
Russia} }

\date{}

\begin{titlepage}
\maketitle
\thispagestyle{empty}

\begin{abstract}
The strong coupling constants between light vector mesons and
octet--decuplet baryons are calculated in framework of the light
cone QCD sum rules, taking into account $SU(3)$ flavor symmetry breaking
effects. It is shown that all strong coupling constants can be
represented in terms of a single universal function. Size of the $SU(3)$
symmetry breaking effects are estimated.
\end{abstract}

%\vspace{1cm}
~~~PACS number(s): 11.55.Hx, 13.30.Eg, 14.20.Jn
\end{titlepage}

\section{Introduction}

In the experiments performed at Jefferson Laboratory, MAMI, BNL, MIT on
meson--nucleon, nucleon--hyperon, hyperon--hyperon reactions
very rich data are accumulated. In order to describe the existing data, the coupling constants of
pseudoscalar and vector mesons with baryons are needed. Calculation of these coupling
constants within QCD, which is the fundamental theory of strong interactions,
constitutes an important problem in studying the dynamics of the
aforementioned reactions.  At the hadronic scale QCD is nonperturbative and
for this reason the calculation of the strong coupling constants of baryons
with vector mesons becomes impossible using fundamental QCD Lagrangian.
Therefore in order to calculate these constants some nonperturbative method
is needed and for this aim we will use QCD sum rules method \cite{R10101},
which is more reliable and predictive in calculating the properties of
hadrons. In the present work, the strong coupling constants of the decuplet--octet
baryons with vector mesons $g$(DOV) are calculated in the framework of light 
cone sum rules method (LCSR) (more about this method and its applications
can be found in \cite{R10102}). Note that, the coupling constants of
pseudoscalar meson octet baryon couplings, vector meson octet baryon
couplings and pseudoscalar meson decuplet baryons are all studied within
LCSR in \cite{R10103}, \cite{R10104} and \cite{R10105}, respectively.
In this work we extend our previous works to investigate the $g$(DOV)
in the LCSR framework.   

The plan of this article is as follows. In section 2 the strong coupling
constants of DOV are calculated in LCSR method and relations between the
above--mentioned strong coupling constants are obtained, when $SU(3)_f$
symmetry breaking effects are taken into account. Section 3 is devoted to
the numerical analysis of the $g$(DOV). 

\section{Light cone QCD sum rules for the vector meson with decuplet--octet
baryon coupling constants}

In $SU(3)_f$ symmetry the coupling constants of all vector mesons with
decuplet--octet baryons are described by the following interaction
Lagrangian,
\bea
\label{nolabel}
{\cal L}_{int} = g \varepsilon_{ijk} \bar{O}_\ell^j D^{mk\ell} V_m^i
+ h.c.~,
\eea
where $O$ $D$ and $V$ are the octet, decuplet baryons and $V$ is the
vector meson, respectively.   

Octet baryons and octet vector mesons are given by,
\bea
\label{nolabel}
O_\beta^\alpha = \left( \begin{array}{ccc}
{1\over \sqrt{2}} \Sigma^0 + {1\over \sqrt{6}} \Lambda & \Sigma^+ & p \\
\Sigma^- & - {1\over 2} \Sigma^0 + {1\over \sqrt{6}} \Lambda  & n \\
\Xi^- & \Xi^0  & -{2\over \sqrt{6}} \Lambda
\end{array} \right)~, \nnb
\eea 
\bea
\label{nolabel}
V_\beta^\alpha = \left( \begin{array}{ccc}
{1\over \sqrt{2}} \rho^0 + {1\over \sqrt{2}} \omega & \rho+ & K^{\ast +} \\
\rho^- & - {1\over \sqrt{2}} \rho^0 + {1\over \sqrt{2}} \omega & K^{\ast 0}
\\
K^{\ast -} & \bar{K}^{\ast 0}  & \phi
\end{array} \right)~. \nnb
\eea

In this section we derive LCSR for the coupling constants of the vector meson
with decuplet--octet baryons. In deriving these sum rules we start our
analysis with the following correlation function,
\bea
\label{e10101}
\Pi_\mu^{D\rar O V} = i \int d^4x e^{ipx} \lla V(q) \vel T\Big\{
\eta (x) \bar{\eta}_\mu (0) \Big\} \ver 0 \rra~,
\eea
where $V(q)$ is the vector meson with momentum $q$, $D$ and $O$ are the
decuplet and octet baryons, $\eta_\mu$ and $\eta$ are their interpolating 
currents respectively, and $T$ is the time ordering operator.

The sum rules for the correlation function (\ref{e10101}) are derived by
calculating it in terms of the hadrons from one side (phenomenological
part), and calculating it in the limit $-p^2 \rar + \infty$ in the deep Euclidean
region in terms of quarks and gluons and then equating these representations
using the dispersion relations.

The phenomenological part can be obtained by inserting a complete set of
intermediate hadronic states with the same quantum numbers as the
corresponding interpolating currents. After isolating the ground state
contributions of the decuplet and octet baryons, we get the following
result,
\bea
\label{e10102}
\Pi_\mu^{D\rar OV} (p,q) = {\lla 0 \vel \eta \ver O(p_2) \rra \lla O(p_2)
V(q) \ve D(p_1) \rra \over p_2^2 - m_2^2 } \;
{\lla D(p_1) \vel \bar{\eta}_\mu \ver 0 \rra \over p_1^2-m_1^2}  + \cdots ~,
\eea
where $O$ and $D$ denote the octet and decuplet baryons,
$p_1=p_2+q$, $p_2$, $m_1$ and $m_2$ are their four--momentum and mass, 
respectively, and $\cdots$ represent the
contributions of the higher states and continuum. The matrix elements in Eq.
(\ref{e10102}) are defined as (see \cite{R10106})
\bea
\label{e10103}
\lla 0 \vel \eta \ver O(p_2) \rra \es \lambda_O u(p_2,s)~, \nnb \\
\lla D(p_1) \vel \bar{\eta}_\mu \ver 0 \rra \es \lambda_D \bar{u}_\mu (p_1,s)~.
\eea
Using the Lorentz invariance, the matrix element $\lla
O(p_2) V(q) \ve D(p_1) \rra$ is parametrized in terms of three form factors
$g_1$, $g_2$ and $g_3$ as follows \cite{R10107}:
\bea
\label{e10104}
\lla O(p_2) V(q) \ve D(p_1) \rra \es \bar{u} (p_2) \Big\{ g_1 (q_\alpha
\rlap/{\varepsilon} - \varepsilon_\alpha \rlap/{q} ) \gamma_5 + 
g_2 [(p\cdot \varepsilon) q_\alpha - (P\cdot q) \varepsilon_\alpha] 
\gamma_5 \nnb \\
\ar g_3 [(q\cdot\varepsilon) q_\alpha - q^2 \varepsilon_\alpha] \gamma_5 \Big\}
u_\alpha (p_1)~,
\eea
where $P=(p_1+p_2)/2$, and $q$ and $\varepsilon$ are the vector meson
momentum and polarization vectors, respectively. In the present work 
we assume the vector meson to be on shell, 
that is $q\cdot \varepsilon = 0$ and $q^2=m_V^2$.

In order to obtain the correlation function from the phenomenological side,
summation over spins of octet and decuplet baryons are performed, i.e.,
\bea
\label{e10105}
\sum_s u(p_2,s) \bar{u}(p_2,s) \es \rlap/{p}_2 + m_2 ~, \nnb \\
\sum_s u_\alpha(p_1,s) \bar{u}_\beta(p_1,s) \es -(\rlap/{p}_1 + m_1)
\Bigg(g_{\alpha\beta} - {\gamma_\alpha \gamma_\beta \over 3 m_1} -
{2 p_{1\alpha} p_{1\beta} \over 3 m_1^2} + {p_{1\alpha} \gamma_\beta -
p_{1\beta} \gamma_\alpha \over 3 m_1} \Bigg) ~.
\eea 
Using Eqs. (\ref{e10102}--\ref{e10105}), in principle, one can obtain the
expression for the phenomenological part of the correlation function.
In doing so, however, the following two problems appear. The first problem
is related to the fact that the interpolating current $\eta_\mu$ couples
not only to the $J^P = {3\over2}^+$ states, but also to the $J^P =
{1\over2}^-$ states. The matrix element of the current $\eta_\mu$ between
vacuum and $J^P = {1\over2}^-$ states is determined by the following
relation
\bea
\label{e10106}
\langle 0 \vel \eta_\mu \ver {1\over 2}^- (p_1) \rangle = A \Bigg( \gamma_\mu - 4
{p_{1\mu} \over m_{{1\over 2}^-}} \Bigg) u(p_1,s)~.
\eea
In deriving this relation the condition $\gamma^\mu p_\mu = 0$ has been
used. Therefore the structures with $\gamma_\mu$ at the end, or the structures
$\sim p_{1\mu}$ contain the contributions from the $J^P = {1\over2}^-$ states
which should be removed, and we will not consider them in further
analysis.

The second problem is that not all Lorentz structures are independent. In
order to overcome these problems we will use the ordering procedure of the
Dirac matrices in such a way that it guarantees the independence of
the Lorentz structures, as well as, it is free of the $J^P = {1\over2}^-$
contributions. In this work we will choose the $\gamma_\mu
\rlap/{\varepsilon}\rlap/{q}\rlap/{p}\gamma_5$ ordering.

Taking into account this ordering and using Eqs.
(\ref{e10102})--(\ref{e10105}) for the phenomenological part of the
correlation function we get
\bea
\label{e10107}
\Pi_\mu \es {\lambda_0 \lambda_D \over [m_1 - (p+q)^2]}
{1\over (m_2^2 -p^2) } \Big[g_1 (m_1+m_2) \rlap/{\varepsilon} \rlap/{p}
\gamma_5 q_\mu - g_2 \rlap/{q} \rlap/{p} \gamma_5 (p\cdot\varepsilon) q_\mu
+ g_3 q^2 \rlap/{q} \rlap/{p} \gamma_5 \varepsilon_\mu \nnb \\
\ar \mbox{\rm other structures} \Big]~.
\eea
 
Obviously, the structures $\rlap/{\varepsilon}\rlap/{p}\gamma_5 q_\mu$,
$\rlap/{q} \rlap/{p} \gamma_5 (p\cdot\varepsilon) q_\mu$ and $\rlap/{q}
\rlap/{p} \gamma_5 \varepsilon_\mu$ do not contain contributions from 
$J^P = {1\over2}^-$ states, because the contributions of $J^P = {1\over2}^-$ 
states are proportional to $\gamma_\mu$ or $(p+q)_\mu$. Note that in 
Eq. (\ref{e10107}) we make the replacements $p_2=p$ and $p_1=p+q$.

For calculation of the theoretical part of the correlation function
from the QCD side the expressions of the interpolating currents of
the decuplet and octet baryons are needed. 

The general form of the interpolating currents for the decuplet baryons
can be presented as \cite{R10106}
\bea
\label{e10108}
\eta_\mu = A \epsilon^{abc} \Big[ \Big( q_1^{aT} C \gamma_\mu q_2^b \Big)
q_3^c + \Big( q_2^{aT} C \gamma_\mu q_3^b \Big) q_1^c +
\Big( q_3^{aT} C \gamma_\mu q_1^b \Big) q_2^c \Big]~,
\eea
where $a,b,c$ are the color indices, $C$ is the charge conjugation operator.
The values of $A$ and the quark flavors $q_1$, $q_2$ and $q_3$ for each
member of the decuplet baryon are listed in Table 1. 

\begin{table}[h]

\renewcommand{\arraystretch}{1.3}
\addtolength{\arraycolsep}{-0.5pt}
\small
$$
\begin{array}{|l|c|c|c|c|}
\hline \hline
 & A & q_1 & q_2 & q_3 \\  \hline
 \Sigma^{\ast 0} & \sqrt{2/3}  & u & d & s  \\
 \Sigma^{\ast +} &            \sqrt{1/3}  & u & u & s  \\
 \Sigma^{\ast -} &            \sqrt{1/3}  & d & d & s  \\
 \Delta^{++}     &        1/3             & u & u & u  \\
 \Delta^{+}      & \sqrt{1/3}  & u & u & d  \\
 \Delta^{0}      & \sqrt{1/3}  & d & d & u  \\
 \Delta^{-}      &        1/3             & d & d & d  \\
 \Xi^{\ast 0}    & \sqrt{1/3}  & s & s & u  \\
 \Xi^{\ast -}    & \sqrt{1/3}  & s & s & d  \\
 \Omega^{-}      &        1/3             & s & s & s  \\
\hline \hline  
\end{array}
$$
\caption{The values of $A$ and the quark flavors $q_1$, $q_2$ and $q_3$}
\renewcommand{\arraystretch}{1}
\addtolength{\arraycolsep}{-1.0pt}

\end{table} 
     
The most general form of the interpolating currents of the octet baryons are
of the following form \cite{R10106,R10108}:
\bea
\label{e10109}
\eta^{\Sigma^0} \es - \sqrt{1\over 2} \epsilon^{abc} \Big[
\Big( u^{aT} C s^b \Big) \gamma_5 d^c - \Big( s^{aT} C d^b \Big)
\gamma_5 u^c +
\beta \Big( u^{aT} C \gamma_5 s^b \Big) d^c - 
\beta \Big( s^{aT} C \gamma_5 d^b \Big) u^c\Big]~, \nnb \\
\eta^{\Sigma^+} \es - \sqrt{1\over 2} \eta^{\Sigma^0}~(d\rar u)~, \nnb \\ 
\eta^{\Sigma^-} \es - \sqrt{1\over 2} \eta^{\Sigma^0}~(u\rar d)~, \nnb \\
\eta^p \es - \eta^{\Sigma^+}~(s\rar d)~, \nnb \\
\eta^n \es - \eta^{\Sigma^-}~(s\rar u)~, \nnb \\
\eta^{\Xi^0} \es - \eta^n ~(d\rar s)~, \nnb \\
\eta^{\Xi^-} \es - \eta^p ~(u\rar s)~, \nnb \\
\eta^\Lambda \es   \sqrt{1\over 6} \epsilon^{abc} \Big[
2 \Big( u^{aT} C d^b \Big) \gamma_5 s^c + \Big( u^{aT} C s^b \Big)
\gamma_5 d^c + \Big( s^{aT} C \gamma_5 d^b \Big) u^c \nnb \\
\ar 2 \beta \Big( u^{aT} C \gamma_5 d^b \Big) s^c + 
\beta \Big( u^{aT} C \gamma_5 s^b \Big) d^c +
\Big( s^{aT} C \gamma_5 d^b \Big) u^c\Big]~, 
\eea
where $\beta$ is an auxiliary parameter and $\beta=-1$ case corresponds to
the Ioffe current. There are the following relations between $\Lambda$ and
$\Sigma^0$ currents, which are shown in 
\cite{R10109},
\bea
\label{e10110}
2 \eta^{\Sigma^0} ~(d \rar s) + \eta^{\Sigma^0} \es - \sqrt{3} \eta^\Lambda~, \nnb \\
2 \eta^{\Sigma^0} ~(u \rar s) + \eta^{\Sigma^0} \es \sqrt{3}
\eta^\Lambda~.
\eea

In principle, having the explicit form of the interpolating currents, the
correlation function can be calculated straightforwardly from the QCD side.
Before calculating the correlation function from the QCD side we will try
to find
the relations among the invariant functions for the structures
$\rlap/{\varepsilon}\rlap/{p}\gamma_5 q_\mu$,
$\rlap/{q} \rlap/{p} \gamma_5 (p\cdot\varepsilon) q_\mu$ and $\rlap/{q}
\rlap/{p} \gamma_5 \varepsilon_\mu$. The relations between the invariant
functions are structure independent, while their explicit terms are structure
dependent. In establishing the relations between invariant functions, we
will follow the works of \cite{R10103,R10104,R10105}. The main power of this
approach, which we present below, is that it it takes into account $SU(3)_f$
symmetry violating effects. We will further show that all correlation
functions which are needed for the determination of the coupling constants
of the vector mesons with decuplet--octet baryons can be written in terms of
only one invariant function for each structure.

Following the works \cite{R10103}--\cite{R10105}, let us
consider the correlation function describing the
$\Sigma^{\ast 0} \rar \Sigma^0 \rho^0$ transition and show that
this transition is described by only one invariant function.
The obtained result is enough to establish relations among
$\Sigma^{\ast 0} \rar \Sigma^0 \rho^0$ and $\Sigma^{\ast +} \rar \Sigma^+
\rho^0$ and $\Sigma^{\ast -} \rar \Sigma^- \rho^0$ transitions. As has
already been noted, the relations among invariant functions are structure
independent, i.e., the relations are the same for all structures. The
correlation function for the $\Sigma^{\ast 0} \rar \Sigma^0 \rho^0$
transition can formally be written as

\bea
\label{e10111}
\Pi^{\Sigma^{\ast 0} \rar \Sigma^0 \rho^0} = g_{\rho^0 uu} \Pi_1(u,d,s) +
g_{\rho^0 dd} \Pi_1^\prime(u,d,s) + g_{\rho^0 ss} \Pi_2(u,d,s)~.
\eea
The meson current is represented as
\bea
\label{nolabel}
J_\mu = \sum_{u,d,s} g_{qq\rho} \bar{q} \gamma_\mu q~, \nnb
\eea
and for the $\rho^0$ meson $g_{\rho^0 uu} = -g_{\rho^0 dd} = 1/\sqrt{2}$ and
$g_{\rho^0 ss}=0$. The functions $\Pi_1$, $\Pi_1^\prime$ and $\Pi_2$ describe
the emission of the $\rho^0$ meson from $u$, $d$ and $s$ quarks,
respectively, and are defined formally as follows:
\bea
\label{e10112}
\Pi_1(u,d,s) \es \lla \bar{u} u \vel \Sigma^0 \bar{\Sigma}^{\ast 0} \ver 0
\rra~, \nnb \\
\Pi_2(u,d,s) \es \lla \bar{s} s \vel \Sigma^0 \bar{\Sigma}^{\ast 0} \ver 0  
\rra~,
\eea

It follows from Eqs. (\ref{e10108}) and (\ref{e10109}) that, the
interpolating currents of $\Sigma^{\ast 0}$ and $\Sigma^0$ are symmetric
with respect to the exchange $u \lrar d$, and for this reason
$\Pi_1^\prime(u,d,s) = \Pi_1(d,u,s)$. Using this result Eq. (\ref{e10111}) 
can be written as
\bea
\label{e10113}
\Pi^{\Sigma^{\ast 0} \rar \Sigma^0 \rho^0} = {1\over \sqrt{2}}
\Big[\Pi_1(u,d,s) - \Pi_1(d,u,s) \Big]~.
\eea
Obviously, in the exact isospin symmetry limit $\Pi^{\Sigma^{\ast 0}
\rar \Sigma^0 \rho^0}=0$.

The invariant functions responsible for $\Sigma^{\ast +} \rar \Sigma^+
\rho^0$ transition can be obtained by replacing $d \rar u$
in $\Pi_1(u,d,s)$ and using $\Sigma^{\ast 0}(d\rar u) = \sqrt{2}
\Sigma^{\ast +}$ and $\Sigma^{0}(d\rar u) = - \sqrt{2} \Sigma^{+}$, which 
gives
\bea
\label{e10114}
- 2 \lla \bar{u} u \vel \Sigma^+ \bar{\Sigma}^{\ast +} \ver 0 \rra = 4
\Pi_1(u,u,s)~.
\eea
The factor four appearing on the right--hand side of Eq. (\ref{e10114}) is
the result of the fact that there are four possible ways that the $\rho^0$ meson 
can be emitted from the $u$ quarks. Using Eq. (\ref{e10114}) we obtain the
following relation for the invariant function $\Pi_1$, which is responsible
for the transition $\Sigma^{\ast +} \rar \Sigma^+ \rho^0$,
\bea
\label{e10115}
\Pi^{\Sigma^{\ast +} \rar \Sigma^+ \rho^0} \es g_{\rho^0 uu} \lla \bar{u} u \vel
\Sigma^+ \bar{\Sigma}^{\ast +} \ver 0 \rra + g_{\rho^0 ss} \lla \bar{s} s \vel 
\Sigma^+ \bar{\Sigma}^{\ast +} \ver 0 \rra \nnb \\
\es - \sqrt{2} \Pi_1(u,u,s)~.
\eea

The relation for the $\Sigma^{\ast -} \rar \Sigma^- \rho^0$ transition can be
obtained by making the replacement $u \rar d$ in Eq. (\ref{e10111}) and
using the fact that $\Sigma^{\ast 0}(u\rar d) = \sqrt{2}
\Sigma^{\ast -}$ and $\Sigma^{0}(u\rar d) = - \sqrt{2} \Sigma^{-}$, 
as a result of which we get,
\bea
\label{e10116}
\Pi^{\Sigma^{\ast -} \rar \Sigma^- \rho^0} = -\sqrt{2}\Pi_1(d,d,s)~.
\eea
Under exact isospin symmetry limit, from Eqs. (\ref{e10115}) and (\ref{e10116})
we get,
\bea
\label{e10117}
\Pi^{\Sigma^{\ast +} \rar \Sigma^+ \rho^0} = \Pi^{\Sigma^{\ast -} \rar
\Sigma^- \rho^0}~.
\eea

We proceed now to calculate the invariant functions involving $\Delta$
resonances. For this aim we consider $\Delta^+ \rar p \rho^0$
transition. Using the fact that $\Delta^+ = \Sigma^{\ast +} (s \rar d)$ and
$p = - \Sigma^+(s \rar d)$, we obtain from Eq. (\ref{e10115}) that,
\bea
\label{e10118}
\Pi^{\Delta^+ \rar p \rho^0} \es - \Big( g_{\rho^0 uu} \lla \bar{u}u \vel \Sigma^+
\bar{\Sigma}^{\ast +} \ver 0 \rra \Big) (s\rar d) - \Big( g_{\rho^0 ss} \lla \bar{s}s
\vel \Sigma^+ \bar{\Sigma}^{\ast +} \ver 0 \rra \Big) (s\rar d) \nnb \\
\es \sqrt{2} \Pi_1(u,u,d) - {1\over \sqrt{2}} \Pi_2(u,u,d)~.
\eea
Obviously, one can easily see that
\bea
\Pi^{\Delta^0 \rar n \rho^0} = \sqrt{2} \Pi_1(d,d,u) - {1\over \sqrt{2}}
\Pi_2(d,d,u)~.
\eea
Following similar lines of reasoning we obtain,
\bea
\label{e10120}
\Pi^{\Xi^{\ast 0} \rar \Xi^0 \rho^0} \es {1\over \sqrt{2}} \Pi_2(s,s,u)~, \nnb \\
\Pi^{\Xi^{\ast -} \rar \Xi^- \rho^0} \es {1\over \sqrt{2}} \Pi_2(s,s,d)~.
\eea

The remaining relations between the correlation functions involving $\rho$, $\omega$ 
and $\phi$ mesons are presented in Appendix A.

Up to this point the relations involving neutral $\rho$ meson are obtained. We can
now try to obtain similar relations among the invariant functions for the
transitions involving charged $\rho$ meson. For this purpose let us 
consider the matrix element $\lla \bar{d}d \vel \Sigma^0
\bar{\Sigma}^{\ast 0} \ver 0 \rra$, where $d$ quarks from $\Sigma^0$ and
$\Sigma^{\ast 0}$ baryons form the final $\bar{d}d$ state, and the remaining
$u$ and $s$ quarks are being the spectators. In the same manner, in the
expression $\lla \bar{u}d \vel \Sigma^+ \bar{\Sigma}^{\ast 0} \ver 0 \rra$,
$d$ quark from $\Sigma^{\ast 0}$ and $u$ quark from $\Sigma^+$ form
$\bar{u}d$ state and the remaining $u$ and $s$ quarks remain as spectators.
Therefore, these matrix elements should be related and the explicit calculations 
indeed confirm this expectation. We find that,
\bea
\label{e10121}
\Pi^{\Sigma^{\ast 0} \rar \Sigma^+ \rho^-} \es \lla \bar{u}d \vel \Sigma^+
\bar{\Sigma}^{\ast 0} \ver 0 \rra = -\sqrt{2} \lla \bar{d}d \vel \Sigma^0
\bar{\Sigma}^{\ast 0} \ver 0 \rra \nnb \\
\es - \sqrt{2} \Pi_1(d,u,s)~.
\eea
Making the exchange $u \lrar d$ in the above expression, we obtain,
\bea
\label{e10122}
\Pi^{\Sigma^{\ast 0} \rar \Sigma^- \rho^+} \es \lla \bar{d}u \vel \Sigma^-  
\bar{\Sigma}^{\ast 0} \ver 0 \rra = \sqrt{2} \lla \bar{u}u \vel \Sigma^0
\bar{\Sigma}^{\ast 0} \ver 0 \rra \nnb \\
\es \sqrt{2} \Pi_1(u,d,s)~.
\eea

Using these arguments and performing similar calculations, we obtain the
following relations among invariant functions involving charged $\rho$
mesons:
\bea
\label{e10123}
\Pi^{\Sigma^{\ast -} \rar \Sigma^0 \rho^-} \es 
\sqrt{2} \Pi_1(u,d,s)~, \nnb \\
\Pi^{\Xi^{\ast -} \rar \Xi^0 \rho^-} \es 
- 2 \Pi_1(d,s,s)~, \nnb \\
\Pi^{\Sigma^{\ast -} \rar \Lambda \rho^-} \es - \sqrt{2\over 3} 
\Big[2 \Pi_1(u,s,d) + \Pi_1(u,d,s) \Big]~, \nnb \\
\Pi^{\Delta^0 \rar p \rho^-} \es 2 \Pi_1(u,u,d)~, \nnb \\
\Pi^{\Sigma^{\ast +} \rar \Sigma^0 \rho^+} \es 
\sqrt{2} \Pi_1(d,u,s)~, \nnb \\     
\Pi^{\Xi^{\ast 0} \rar \Xi^- \rho^+} \es
- \Pi_2(s,s,u)~, \nnb \\
\Pi^{\Sigma^{\ast +} \rar \Lambda \rho^+} \es 
\sqrt{2\over 3} \Big[2 \Pi_1(d,s,u) + \Pi_1(d,u,s) \Big]~, \nnb \\
\Pi^{\Delta^+ \rar n \rho^+} \es - 2 \Pi_1(d,d,u)~.
\eea
The relations among the invariant functions involving $K^\ast$ and $\phi$
mesons can be obtained easily using the similar arguments. These relations
are presented in Appendix A for a given Lorentz structure. 

All of the obtained results for the coupling constants of DOV can
qualitatively be understood from a simple {\it diquark+quark} picture of the
baryons in the following way. Let us consider the Hermitian conjugate
channel, i.e., $O+V\rar D$. Denote $O$ as $O(q_1 q_2,q_3)$ where 
$q_1 q_2$ forms the diquark and $q_3$ is a single quark. In a reaction,
for example $p \rho^+ \rar \Delta^{++}$, the single
$q_3$ (in this case $d$) is the $V$--absorbing quark, and therefore, this reaction
is described by $\Pi_{q_3}$ (in our notation $\Pi_2$). 
In the reaction $n \rho^+ \rar \Delta^+$, the
$V$--absorbing quark is from the diquark (here $dd$), and this reaction is
described by, let us say, $\Pi_1$. 

For the reactions of $\Delta$ decays with the participation of $\rho^0$,
both functions $\Pi_1$ and $\Pi_2$ contribute. Obviously, only $\Pi_1$ and
$\Pi_2$ contribute to the reactions $\Sigma^\ast \rar \Sigma \rho^0$ and 
$\Xi^\ast \rar \Xi \rho^0$, respectively. Similar situation take place for
the reactions involving $\omega$ and $\phi$ mesons. 

From calculations the following relation between $\Pi_1$ and $\Pi_2$ is 
obtained
\bea
\label{nolabel}
\Pi_2 (u,d,s) = - \Pi_1 (s,u,d) - \Pi_1 (s,d,u)~.\nnb
\eea
This relation leads to the result that
all coupling constants of the vector mesons with the decuplet--octet baryons
can be written in terms of only one invariant function without using
the $SU(3)_f$ flavor symmetry, which is the main result of this work.

We now concentrate on calculating the invariant function $\Pi_1$. 
For this aim, the correlation function which describes
the transition $\Sigma^{\ast 0} \rar \Sigma^0 \rho^0$ is enough.
In deep Euclidean region, $-p_1^2 \rar \infty$, $-p_2^2 \rar \infty$, as we
have already mentioned, the
correlation function can be evaluated from the QCD side using OPE. 
In order to obtain the expressions of the correlation functions from QCD side, 
the propagator of the light quarks and the matrix elements of the nonlocal 
operators $\bar{q}(x_1) \Gamma q^\prime (x_2)$ and $\bar{q}(x_1) G_{\mu\nu} 
q^\prime (x_2)$ between the vacuum and the vector meson states are needed,
where $\Gamma$ represents the Dirac matrices relevant to the case under
consideration, and $G_{\mu\nu}$ is the gluon field strength tensor. 

Up to twist--4 accuracy,
matrix elements $\lla V(q) \vel \bar{q}(x) \Gamma q(0) \ver 0 \rra$ and 
$\lla V(q) \vel \bar{q}(x) G_{\mu\nu} q(0) \ver 0 \rra$ are determined in
terms of the distribution amplitudes (DA's) of the vector mesons
\cite{R10110,R10111,R10112}. These DA's are presented in
Appendix B.

In further analysis, we use the following
expression for the light quark propagator
\bea
\label{e10125}
S_q(x) \es {i \rlap/x \over 2 \pi^2 x^4} - {m_q \over 4 \pi^2 x^2} -
{\lla\bar{q}q\rra \over 12} \Bigg(1 - {i m_q \over 4} \rlap/x \Bigg) -
{x^2 \over 192} m_0^2 \lla\bar{q}q\rra  \Bigg(1 - {i m_q \over 6} 
\rlap/x \Bigg) \nnb \\
\ek i g_s \int_0^1 du \Bigg\{ {\rlap/x \over 16 \pi^2 x^2} G_{\mu\nu} (ux)
\sigma^{\mu\nu} - u x^\mu G_{\mu\nu} (ux) \gamma^\nu {i \over 4 \pi^2 x^2}
\nnb \\
\ek {i m_q \over 32 \pi^2} G_{\mu\nu} (ux) \sigma^{\mu\nu} \Bigg[ \ln \Bigg(
{- x^2 \Lambda^2 \over 4} \Bigg) + 2 \gamma_E \Bigg] \Bigg\}~,
\eea 
where $\gamma_E$ is the Euler constant, $\Lambda$ is a scale parameter, and
we will choose it as a factorization scale, i.e., $\Lambda = 1.0~GeV$ 
(for more detail, see \cite{R10113}). In the calculations, $SU(3)_f$
symmetry violation effects are included in the nonzero strange quark mass and
strange quark condensate. These effects are also taken into
account in calculation of the DA's \cite{R10110,R10111,R10112}.

Having the expressions of the light quark propagator and those of
the DA's, the theoretical part of the correlation 
functions can be calculated. Equating both representations of correlation
function and separating coefficients of Lorentz structures 
$\rlap/{\varepsilon}\rlap/{p}\gamma_5 q_\mu$,
$\rlap/{q} \rlap/{p} \gamma_5 (p\cdot\varepsilon) q_\mu$ and $\rlap/{q}
\rlap/{p} \gamma_5 \varepsilon_\mu$, and
applying Borel transformation to both side of the correlation functions
on the variables $p^2$ and $(p+q)^2$ in order to suppress the
contributions of the higher states and continuum (see
\cite{R10114}), we get the sum rules for the corresponding vector
meson decuplet--octet baryon couplings. The contributions of higher states and 
the continuum are subtracted using quark-hadron duality.
After standard calculations, for each Lorentz
structure the expressions for the invariant functions 
$\Pi_1^{(\alpha)}$ are obtained and their expressions are
presented in Appendix C.
Here superscript $\alpha$ refers to the invariant functions
$\Pi^{(\alpha)}$ relevant to the coupling constants $g_1$, $g_2$ and $g_3$,
correspondingly.

For a given transition $D \rightarrow O V$, once the Borel transformed and 
continuum subtracted 
coefficient functions $\Pi^{(1)}$, $\Pi^{(2)}$ and $\Pi^{(3)}$ are obtained,
the coupling constants can be written as
\bea
\label{e10126}
g_1 \es {1\over m_1+m_2} {1\over \lambda_{O} \lambda_{D}}e^{{m_1^2 \over M_1^2} + 
{m_2^2 \over M_2^2} + {m_V^2 \over{M_1^2 + M_2^2}}} \, \Pi_1^{(1)}~,
\nnb \\
g_2 \es {1\over \lambda_{O} \lambda_{D}}e^{{m_1^2 \over M_1^2} + 
{m_2^2 \over M_2^2} + {m_V^2 \over{M_1^2 + M_2^2}}} \, \Pi_1^{(2)}~, \nnb \\
g_3 \es {1\over m_V^2} {1\over \lambda_{O} \lambda_{D}}e^{{m_1^2 \over M_1^2} + 
{m_2^2 \over M_2^2} + {m_V^2 \over{M_1^2 + M_2^2}}} \, \Pi_1^{(3)}~.
\eea
It follows from these expressions that for the vector meson decuplet--octet 
baryon strong coupling constants, the
residues of baryons are needed. The residues of baryons are obtained from
the analysis of two--point correlation function as are given in
\cite{R10106,R10108,R10114}. The currents of the other baryons can be obtained from
$\Sigma^0$ current by making appropriate substitutions of quarks. For this
reason, for determination of the residues, we give the sum rule only for
$\Sigma^0$ and $\Sigma^{\ast 0}$.
\bea
\label{e10127}
\lambda_{\Sigma^0}^2 e^{-m_{\Sigma^0}^2/M^2} \es
{M^6\over 1024 \pi^2} (5 + 2 \beta + 5 \beta^2) E_2(x) - {m_0^2\over 96 M^2} (-1+\beta)^2 \lla
\bar{u} u \rra \lla \bar{d} d \rra \nnb \\
\ek {m_0^2\over 16 M^2} (-1+\beta^2) \lla \bar{s} s \rra \Big(\lla \bar{u} u \rra +
\lla \bar{d} d \rra\Big) \nnb \\
\ar {3 m_0^2\over 128} (-1+\beta^2) \Big[ m_s \Big(\lla \bar{u} u \rra +   
\lla \bar{d} d \rra\Big) + (m_u+m_d) \lla \bar{s} s \rra \Big] \nnb \\
\ek  {1\over 64 \pi^2} (-1+\beta)^2 M^2 \Big( m_d \lla \bar{u} u \rra + m_u \lla \bar{d} d
\rra \Big) E_0(x) \nnb \\
\ek {3 M^2\over 64 \pi^2} (-1+\beta^2) \Big[ m_s \Big(\lla \bar{u} u \rra +   
\lla \bar{d} d \rra\Big) + (m_u+m_d) \lla \bar{s} s \rra \Big] E_0(x) \nnb \\
\ar {1\over 128 \pi^2}  (5 + 2 \beta + 5 \beta^2) \Big( m_u \lla \bar{u} u \rra +
m_d \lla \bar{d} d \rra + m_s \lla \bar{s} s \rra\Big) \nnb \\
\ar {1\over 24} \Big[ 3 (-1+\beta^2) \lla \bar{s} s \rra \Big(\lla \bar{u} u
\rra + \lla \bar{d} d \rra \Big) + (-1+\beta^2) \lla \bar{u} u \rra \lla \bar{d}
d \rra \Big] \nnb \\
\ar {m_0^2\over 256 \pi^2} (-1+\beta)^2 \Big(m_u \lla \bar{d} d \rra +   
m_d \lla \bar{u} u \rra\Big) \nnb \\
\ar {m_0^2\over 26 \pi^2} (-1+\beta^2) \Big[ 13 m_s \Big(\lla \bar{u} u \rra +   
\lla \bar{d} d \rra\Big) + 11 (m_u+m_d) \lla \bar{s} s \rra \Big] \nnb \\
\ek {m_0^2\over 192 \pi^2}(1+\beta+\beta^2) \Big( m_u \lla \bar{u} u \rra +
m_d \lla \bar{d} d \rra - 2 m_s \lla \bar{s} s \rra\Big)~, \nnb \\ \nnb \\  
m_{\Sigma^{\ast 0}} \lambda_{\Sigma^{\ast 0}}^2 e^{-{m_\Sigma^2\over M^2}} \es
\left( \uu + \dd + \sp \right) {M^4\over 9 \pi^2} E_1(x)
- \left( m_u + m_d + m_s\right) {M^6\over 32 \pi^4} E_2(x) \nnb \\
\ek \left( \uu + \dd + \sp \right) m_0^2 {M^2\over 18 \pi^2} E_0(x) \nnb \\
\ek {2\over 3}\left(1 + {5 m_0^2\over 72 M^2} \right) \left( m_u \dd \sp +
m_d \sp \uu + m_s \dd \uu \right) \nnb \\
\ar \left( m_s \dd \sp + m_u \dd \uu + m_d \sp \uu \right) {m_0^2\over 12
M^2}~,
\eea
where $x = s_0/M^2$, and
\bea
\label{nolabel}
E_n(x)=1-e^{-x}\sum_{i=0}^{n}\frac{x^i}{i!}~. \nnb
\eea

The contribution of the higher states and continuum in the invariant
functions are subtracted by taking
into account  the following replacements

\bea
\label{e10128}
e^{-m_V^2/4 M^2} M^2 \left( \ln{M^2 \over \Lambda^2} - \gamma_E \right)
\erar \int_{m_V^2/4}^{s_0}  ds  e^{- s/M^2}
\ln {s-m_V^2/4\over \Lambda^2}\nnb \\
e^{-m_V^2/4 M^2} \left( \ln{M^2 \over \Lambda^2} - \gamma_E \right)
\erar \ln {s_0 - m_V^2/4\over \Lambda^2} e^{-s_0/M^2} +
{1\over M^2} \int_{m_V^2/4}^{s_0} ds e^{- s/M^2}
\ln {s-m_V^2/4\over \Lambda^2}\nnb \\
e^{-m_V^2/4 M^2} {1\over M^2} \left( \ln{M^2 \over \Lambda^2} -
\gamma_E \right)
\erar {1\over M^2} \ln
{s_0 - m_V^2/4\over \Lambda^2} e^{-s_0/M^2}
+ {1\over s_0-m_V^2/4}  e^{-s_0/M^2} \nnb \\
\ar {1\over M^4} \int_{m_V^2/4}^{s_0} ds e^{- s/M^2}
\ln {s-m_V^2/4\over \Lambda^2}\nnb \\
e^{-m_V^2 / 4 M^2} M^{2n}
\erar {1\over \Gamma(n)}
\int_{m_V^2/4}^{s_0} ds  e^{- s/M^2} \left( s - m_V^2/4
\right)^{n-1}~.
\eea

\section{Numerical analysis and discussion}

This section is devoted to the numerical analysis of the sum rules for the
vector meson octet--decuplet baryon coupling constants. The main input 
parameters of the light cone sum rules in our case are the vector meson 
DA's. The DA's of the vector mesons are given in    
\cite{R10110,R10111,R10112}. The values of the leptonic constants $f_V$ and 
$f_V^T$, and of the twist--2 and twist--3 parameters $a_i^\parallel$,
$a_i^\perp$,$\zeta_{3V}^\parallel$, $\tilde{\lambda}_{3V}^\parallel$,
$\tilde{\omega}_{3V}^\parallel$, $\kappa_{3V}^\parallel$,
$\omega_{3V}^\parallel$, $\lambda_{3V}^\parallel$, $\kappa_{3V}^\perp$,
$\omega_{3V}^\perp$, $\lambda_{3V}^\perp$, as well as twist--4 parameters 
$\zeta_4^\parallel$, $\tilde{\omega}_4^\parallel$, $\zeta_4^\perp$,
 $\tilde{\zeta}_4^\perp$, $\kappa_{4V}^\parallel$, $\kappa_{4V}^\perp$ are
given in Table (1) and Table (2), respectively, in \cite{R10112}.
The value of the other input parameters which are needed in the sum rule are
$\lla \bar{q} q \rra = - (0.243~GeV)^3$, $m_0^2 = 0.8$ \cite{R10108},
$\GG = 0.47~GeV^4$ \cite{R10101}.  

It should be noted here that, the masses of initial and final baryons
are close to each other. Therefore we can choose
$M_1^2=M_2^2=2 M^2$, and consequently $u_0=1/2$. Hence, in further
numerical analysis, the values of the DA's only at $u_0=1/2$ are needed.  
  
It follows from the explicit expressions of the sum rules for the vector 
meson decuplet--octet baryon coupling constants that, in addition to the DA's, 
they also contain three auxiliary parameters, namely, Borel mass parameter,
continuum threshold $s_0$, and the parameter $\beta$ in the interpolating 
current. Therefore, we need to find the working regions of those parameters 
where the results of the vector meson decuplet--octet baryon coupling 
constants are practically independent of them.

The upper bound of the Borel parameter $M^2$ can be found  by requiring
that the higher states and continuum contributions to a correlation
function should be less than, say 40\% of the total value of the same correlation
function. The lower bound of $M^2$ can be found by requiring that the
contribution of the highest term with the power of $1/M^2$ be 20--25\% less than
that of the highest power of $M^2$. Using these restrictions, we obtain the working 
region for the Borel parameters. The continuum threshold is varied in the regions
$s_0=(m_B+0.5)^2$ and $s_0=(m_B+0.7)^2$.
     
In Fig. (1)--(3) we present the dependence of the couplings $g_1$, $g_2$ and
$g_3$ for the $\Sigma^{\ast +} \rar \Sigma^+ \rho^0$ transition at five 
different values of the parameter $\beta$ at $s_0=4~GeV^2$. We observe from 
these figures that these couplings have good stability in the ``working" region 
of $M^2$. Obviously, the coupling constants are also expected to be independent
of the auxiliary parameter $\beta$. In order to find the working region of $\beta$
where these couplings, we depict in Figs. (4)--(5) the dependence of $g_1$,
$g_2$ and $g_3$ on $\cos \theta$ for the $\Sigma^{\ast +} \rar \Sigma^+
\rho^0$ transition, where $\theta$ is defined as $\tan\theta = \beta$. It
follows from these figures that the common working region of $\beta$ for the
coupling constants is $-0.5 < \cos\theta < 0.3$, where these constants exhibit
weak dependence on $\beta$. Up to now we apply the standard procedure in
analysis of the sum rules, i.e., $s_0$ is chosen to be independent of $M^2$
and $q^2$. However, in general the continuum threshold
should depend on $M^2$ and momentum transfer
squared $q^2$ \cite{R10115}. This dependence leads to, so called,
systematic uncertainties, and for this reason, the usual criteria 
for the stability of the results on $M^2$ does not provide
ordering of the realistic errors. We consider these
systematic errors to be about 15\% (see \cite{R10115}). 
Under these conditions, the result of our
analysis predict that $g_1=6 \pm 2$, $g_2=2 \pm 0.5$ and $g_3=20 \pm 4$.

The same coupling constants for the other transitions
are presented in Table 1 (here we present results only for the absolute values of
the coupling constants). Only those results which could not be obtained by
the $SU(2)$ symmetry rotations are presented in this table. 
The errors presented in Table--1 take into account the uncertainties coming from
the variation of $M^2$, $s_0$ and $\beta$, as well as uncertainties inherit
in the input parameters and systematic errors.

From Tables 1--3 we get the following results:

\begin{itemize}

\item As far as the coupling constant $g_1$ is concerned, for the channels
presented in Table--1, we observe that there is good agreement between the
predictions of the  general current and the Ioffe current for the octet
baryons. All of the results
presented within the limit of errors satisfy the relations among the
coupling constants and invariant functions.

\item In the case of the coupling constant $g_2$, there are considerable
discrepancies between the predictions of the general current and Ioffe
current for the octet baryons for the $\Sigma^{\ast +} \rar \Xi^0 K^{\ast
+}$, $\Omega^- \rar \Xi^0 K^{\ast -}$, $\Delta^+ \rar \Sigma^0 K^{\ast +}$,
$\Delta^{++} \rar \Sigma^+ K^{\ast +}$, $\Xi^{\ast 0} \rar \Lambda
\bar{K}^{\ast 0}$, $\Delta^- \rar \Sigma^- \bar{K}^{\ast 0}$ 
and $\Xi^{\ast 0} \rar \Sigma^+ K^{\ast -}$ channels. 
The characteristic property of all these
channels is that they all involve the $K^\ast$ meson. Note that, for
these channels, the relations among the coupling constants and invariant
functions are also strongly violated. In the general current case all these
discrepancies can be attributed to not having stable region for
$\cos\theta$. In other words the results for the coupling constants of these
transitions are not reliable. For the remaining transitions the results for
both cases are close to each other within the error limits.

\item For the coupling constant $g_3$ our results can be summarized as
follows: The relations between the coupling constant $g_3$ and the invariant
function $\Pi_1$ are strongly violated in the channels $\Sigma^{\ast -} \rar
\Lambda \rho^-$, $\Xi^{\ast 0} \rar \Xi^0 \rho^0$, as well as in the
channels involving $\phi$ meson.
These discrepancies, as is the case for the $g_2$ coupling constant, can be
attributed to the absence of the stability region $g_3$ with respect to the
$\cos\theta$. For this reason, for the above--mentioned transitions, the 
predictions for the coupling constant $g_3$ are not reliable.

\end{itemize}

The gap between the results of the two currents can be attributed to the
fact that, for many transitions the value $\beta=-1$ lies outside the stability 
region of $\beta$, causing considerable discrepancies between the
predictions of the two currents. 

In conclusion, the strong coupling constant of the vector mesons with
decuplet--octet baryons are studied within LQSR. It is shown that all
coupling constants can be calculated in terms of a single universal function.

\newpage 
     
% .........................................................

\begin{table}[h]

\renewcommand{\arraystretch}{1.3}
\addtolength{\arraycolsep}{-0.5pt}
\small
$$
\begin{array}{|l|r@{\pm}l|c|r@{\pm}l|c|}
\hline \hline  
 \multirow{2}{*}{$g_1^{\mbox{\small{\,channel}}}$} &  
 \multicolumn{3}{c|}{\mbox{General current}}       & 
 \multicolumn{3}{c|}{\mbox{Ioffe current}}           \\
&    \multicolumn{2}{c}{\mbox{Result}}  &  {\mbox{$SU(3)_f$}} 
&    \multicolumn{2}{c}{\mbox{Result}}  &  {\mbox{$SU(3)_f$}} \\ \hline
 g_1^{\Sigma^{\ast +} \rar \Sigma^+ \rho^0}        &  -5.6&1.6  &   -5.7  &   -7.8&0.5 & -7.8  \\ 
 g_1^{\Delta^0 \rar p \rho^-}                      &   9.1&2.9  &    8    &   12.2&0.8 &  11   \\
 g_1^{\Xi^{\ast 0} \rar \Xi^0 \rho^0}              &  -5.5&1.5  &   -5.7  &   -7.2&0.3 & -7.8  \\
 g_1^{\Sigma^{\ast -} \rar \Lambda \rho^-}         &  - 10&3    &   -9.8  &  -13.5&0.7 & -13.5 \\
 g_1^{\Delta^+ \rar \Sigma^0 K^{\ast +}}           & -13.1&4.1  &  -11.3  &  -18.4&1.4 & -15.6 \\
 g_1^{\Sigma^{\ast +} \rar \Xi^0 K^{\ast +}}       &   8.9&2.5  &    8    &     12&0.5 &  11   \\
 g_1^{\Sigma^{\ast +} \rar p \bar{K}^{\ast 0}}     &  -9.5&2.7  &   -8    &  -12.4&0.9 & -11   \\
 g_1^{\Omega^- \rar \Xi^0 K^{\ast -}}              &  - 14&4    &  -13.9  &  -18.8&1.4 & -19.1 \\
 g_1^{\Xi^{\ast 0} \rar \Sigma^+ K^{\ast -}}       &  -8.6&2.5  &   -8    &  -11.6&0.9 & -11   \\
 g_1^{\Xi^{\ast 0} \rar \Lambda \bar{K}^{\ast 0}}  &  10.8&3.0  &    9.8  &   14.1&0.9 &  13.5 \\
 g_1^{\Sigma^{\ast +} \rar \Sigma^+ \omega}        &  -5.1&1.4  &   -5.7  &   -6.9&0.4 & -7.8  \\
 g_1^{\Xi^{\ast 0} \rar \Xi^0 \omega}              &  -4.9&1.3  &   -5.7  &   -6.5&0.5 & -7.8  \\
 g_1^{\Sigma^{\ast +} \rar \Sigma^+ \phi}          &   8.0&2.8  &    8    &   11.6&0.7 &  11   \\
 g_1^{\Xi^{\ast 0} \rar \Xi^0 \phi}                &   7.1&1.9  &    8    &   10.9&0.8 &  11   \\
 \hline \hline
\end{array}
$$
\caption{The values of the coupling constant $g_1$ for various channels.}
\renewcommand{\arraystretch}{1}
\addtolength{\arraycolsep}{-1.0pt}

\end{table}

\newpage

% .........................................................

\begin{table}[h]

\renewcommand{\arraystretch}{1.3}
\addtolength{\arraycolsep}{-0.5pt}
\small
$$
\begin{array}{|l|r@{\pm}l|c|r@{\pm}l|c|}
\hline \hline  
 \multirow{2}{*}{$g_2^{\mbox{\small{\,channel}}}$} &  
 \multicolumn{3}{c|}{\mbox{General current}}       & 
 \multicolumn{3}{c|}{\mbox{Ioffe current}}           \\
&    \multicolumn{2}{c}{\mbox{Result}}  &  {\mbox{$SU(3)_f$}} 
&    \multicolumn{2}{c}{\mbox{Result}}  &  {\mbox{$SU(3)_f$}} \\ \hline
 g_2^{\Sigma^{\ast +} \rar \Sigma^+ \rho^0}        &  -2.3&1    &   -2.3  &   -3.6&0.6 & -3.8  \\ 
 g_2^{\Delta^0 \rar p \rho^-}                      &   3.3&1.5  &    3.2  &    5.3&1.0 &  5.4  \\
 g_2^{\Xi^{\ast 0} \rar \Xi^0 \rho^0}              &  -2.2&1    &   -2.3  &   -3.4&0.5 & -3.8  \\
 g_2^{\Sigma^{\ast -} \rar \Lambda \rho^-}         &    -4&1.7  &   -3.9  &   -6.3&1.2 & -6.6  \\
 g_2^{\Delta^+ \rar \Sigma^0 K^{\ast +}}           & -16.4&6    &   -4.5  &  -27.3&7.3 & -7.6  \\
 g_2^{\Sigma^{\ast +} \rar \Xi^0 K^{\ast +}}       &  11.1&3    &    3.2  &   16.9&3.5 &  5.4  \\
 g_2^{\Sigma^{\ast +} \rar p \bar{K}^{\ast 0}}     & -10.8&4.8  &   -3.2  &  -17.9&4.7 & -5.4  \\
 g_2^{\Omega^- \rar \Xi^0 K^{\ast -}}              & -17.3&6    &   -5.5  &  -27.7&6.0 & -9.4  \\
 g_2^{\Xi^{\ast 0} \rar \Sigma^+ K^{\ast -}}       &  -6.3&4.0  &   -3.2  &  -16.1&3.4 & -5.4  \\
 g_2^{\Xi^{\ast 0} \rar \Lambda \bar{K}^{\ast 0}}  &  12.4&4.8  &    3.9  &   19.2&3.9 &  6.6  \\
 g_2^{\Sigma^{\ast +} \rar \Sigma^+ \omega}        &  -2.0&0.7  &   -2.3  &   -3.2&0.6 & -3.8  \\
 g_2^{\Xi^{\ast 0} \rar \Xi^0 \omega}              &  -1.9&0.6  &   -2.3  &   -3.0&0.4 & -3.8  \\
 g_2^{\Sigma^{\ast +} \rar \Sigma^+ \phi}          &   4.4&1.8  &    3.2  &    7.2&1.5 &  5.4  \\
 g_2^{\Xi^{\ast 0} \rar \Xi^0 \phi}                &   4.2&2    &    3.2  &    6.7&1.1 &  5.4  \\
 \hline \hline
\end{array}
$$
\caption{The values of the coupling constant $g_2$ for various channels.}
\renewcommand{\arraystretch}{1}
\addtolength{\arraycolsep}{-1.0pt}

\end{table}

\newpage
     
% .........................................................

\begin{table}[h]

\renewcommand{\arraystretch}{1.3}
\addtolength{\arraycolsep}{-0.5pt}
\small
$$
\begin{array}{|l|r@{\pm}l|c|r@{\pm}l|c|}
\hline \hline  
 \multirow{2}{*}{$g_3^{\mbox{\small{\,channel}}}$} &  
 \multicolumn{3}{c|}{\mbox{General current}}       & 
 \multicolumn{3}{c|}{\mbox{Ioffe current}}           \\
&    \multicolumn{2}{c}{\mbox{Result}}  &  {\mbox{$SU(3)_f$}} 
&    \multicolumn{2}{c}{\mbox{Result}}  &  {\mbox{$SU(3)_f$}} \\ \hline
 g_3^{\Sigma^{\ast +} \rar \Sigma^+ \rho^0}        &  31.4&3.5  &   12.9   &  27.4&1.8 &  18.7  \\ 
 g_3^{\Delta^0 \rar p \rho^-}                      & -33.5&3.1  &  -18.2   & -35.2&2.6 & -26.4  \\
 g_3^{\Xi^{\ast 0} \rar \Xi^0 \rho^0}              &  16.2&3.8  &   12.9   &  27.9&1.6 &  18.7  \\
 g_3^{\Sigma^{\ast -} \rar \Lambda \rho^-}         &  36.5&8.5  &   22.3   &  45.7&2.6 &  32.3  \\
 g_3^{\Delta^+ \rar \Sigma^0 K^{\ast +}}           &  22.6&2.4  &   25.7   &  19.6&1.9 &  37.3  \\
 g_3^{\Sigma^{\ast +} \rar \Xi^0 K^{\ast +}}       & -17.5&2.3  &  -18.2   & -17.3&1.8 & -26.4  \\
 g_3^{\Sigma^{\ast +} \rar p \bar{K}^{\ast 0}}     &  14.6&1.7  &   18.2   &  12.6&2.6 &  26.4  \\
 g_3^{\Omega^- \rar \Xi^0 K^{\ast -}}              &  33.5&3    &   31.5   &  31.4&2.4 &  45.7  \\
 g_3^{\Xi^{\ast 0} \rar \Sigma^+ K^{\ast -}}       &  17.3&0.7  &   18.2   &  16.5&1.4 &  26.4  \\
 g_3^{\Xi^{\ast 0} \rar \Lambda \bar{K}^{\ast 0}}  & -36.5&8.2  &  -22.3   & -45.5&1.7 & -32.3  \\
 g_3^{\Sigma^{\ast +} \rar \Sigma^+ \omega}        &  23.9&6.5  &   12.9   &  27.9&2.1 &  18.7  \\
 g_3^{\Xi^{\ast 0} \rar \Xi^0 \omega}              &  19.1&4.3  &   12.9   &  23.9&1.3 &  18.7  \\
 g_3^{\Sigma^{\ast +} \rar \Sigma^+ \phi}          & -24.2&6.2  &  -18.2   & -31.7&2.1 & -26.4  \\
 g_3^{\Xi^{\ast 0} \rar \Xi^0 \phi}                & -16.2&3.8  &  -18.2   & -33.2&1.7 & -26.4  \\
 \hline \hline
\end{array}
$$
\caption{The values of the coupling constant $g_3$ for various channels.}
\renewcommand{\arraystretch}{1}
\addtolength{\arraycolsep}{-1.0pt}

\end{table}

\newpage

\bAPP{A}{}

In this appendix we present the relations among the correlation functions
involving $\rho$, $\omega$, $\phi$ and $K^\ast$ mesons.

\begin{itemize}
\item The vertices involving $\rho$ meson.
\end{itemize}

\baeeq
\label{e101ap01}
\Pi^{\Sigma^{\ast 0} \rar \Lambda \rho^0} \es - {1\over \sqrt{6}} 
[2 \Pi_1(u,s,d) + \Pi_1(u,d,s) - \Pi_1(d,u,s) - 
2 \Pi_1(d,s,u) ]~, \nnb \\
\Pi^{\Delta^- \rar n \rho^-} \es 2 \sqrt{3} \Pi_1(d,d,d)~, \nnb \\
\Pi^{\Delta^{++} \rar p \rho^+} \es - 2 \sqrt{3} \Pi_1(u,u,u)~.
\eaeeq

\begin{itemize}
\item The vertices involving $K^\ast$ meson.
\end{itemize}
 
\baeeq
\label{e101ap02}
\Pi^{\Delta^+ \rar \Sigma^0 K^{\ast +}} \es -\sqrt{2} [
 \Pi_1(s,u,d) - \Pi_1(s,d,u)]~, \nnb \\
\Pi^{\Delta^+ \rar \Lambda K^{\ast +}} \es {\sqrt{2}\over \sqrt{3}}
[\Pi_1(s,d,u) - \Pi_1(s,u,d) ]~, \nnb \\
\Pi^{\Delta^0 \rar \Sigma^- K^{\ast +}} \es -2 \Pi_1(s,d,d)~, \nnb \\
\Pi^{\Sigma^{\ast +} \rar \Xi^0 K^{\ast +}} \es 2 \Pi_1(u,s,u )~, \nnb \\
\Pi^{\Sigma^{\ast 0} \rar \Xi^- K^{\ast +}} \es \sqrt{2} \Pi_1(u,s,d )~, \nnb \\
\Pi^{\Delta^{++} \rar \Sigma^+ K^{\ast +}} \es - 2 \sqrt{3} \Pi_1(s,u,u)~, \nnb \\
\Pi^{\Sigma^{\ast 0} \rar p K^{\ast -}} \es \sqrt{2} \Pi_1(u,u,d )~, \nnb \\
\Pi^{\Omega^- \rar \Xi^0 K^{\ast -}} \es - \sqrt{2} \Pi_1(u,d,s )~, \nnb \\
\Pi^{\Sigma^{\ast -} \rar n K^{\ast -}} \es 2 \sqrt{2} \Pi_1(s,d,d)~, \nnb \\
\Pi^{\Xi^{\ast 0} \rar \Sigma^+ K^{\ast -}} \es - 2 \Pi_1(s,d,d)~, \nnb \\
\Pi^{\Xi^{\ast -} \rar \Sigma^0 K^{\ast -}} \es \sqrt{2} \Pi_1(u,d,s )~, \nnb \\
\Pi^{\Xi^{\ast -} \rar \Lambda K^{\ast -}} \es - {\sqrt{2}\over \sqrt{3}}
[2 \Pi_1(u,s,d) + \Pi_1(u,d,s) ]~, \nnb \\
\Pi^{\Xi^{\ast 0} \rar \Sigma^0 K^{\ast 0}} \es \sqrt{2} \Pi_1(d,u,s )~, \nnb \\
\Pi^{\Xi^{\ast 0} \rar \Lambda K^{\ast 0}} \es {\sqrt{2}\over \sqrt{3}}
[2 \Pi_1(d,s,u) + \Pi_1(d,u,s) ]~, \nnb \\
\Pi^{\Xi^{\ast -} \rar \Sigma^- K^{\ast 0}} \es 2 \Pi_1(d,d,s )~, \nnb \\
\Pi^{\Sigma^{\ast 0} \rar n K^{\ast 0}} \es - \sqrt{2} \Pi_1(d,d,u )~, \nnb \\
\Pi^{\Omega^- \rar \Xi^- K^{\ast 0}} \es 2 \sqrt{3} \Pi_1(s,s,s)~, \nnb \\
\Pi^{\Sigma^{\ast +} \rar p K^{\ast 0}} \es - 2 \Pi_1(s,u,u)~, \nnb \\
\Pi^{\Sigma^{\ast 0} \rar \Xi^0 \bar{K}^{\ast 0}} \es \sqrt{2} \Pi_1(d,s,u )~, \nnb \\
\Pi^{\Delta^- \rar \Sigma^- \bar{K}^{\ast 0}} \es - 2 \sqrt{3} \Pi_1(d,d,d)~, \nnb \\
\Pi^{\Sigma^{\ast -} \rar \Xi^- \bar{K}^{\ast 0}} \es - 2 \Pi_1(d,s,d )~, \nnb \\
\Pi^{\Delta^0 \rar \Sigma^0 \bar{K}^{\ast 0}} \es - \sqrt{2} [ 
\Pi_1(s,d,u) + \Pi_1(s,u,d) ]~, \nnb \\
\Pi^{\Delta^0 \rar \Lambda \bar{K}^{\ast 0}} \es {\sqrt{2} \over \sqrt{3}}
[ \Pi_1(s,d,u) - \Pi_1(s,u,d) ]~, \nnb \\
\Pi^{\Delta^+ \rar \Sigma^+ \bar{K}^{\ast 0}} \es 2 \Pi_1(s,u,u)~.
\eaeeq

\begin{itemize}
\item Finally, we present the expressions of the vertices involving $\omega$ and
$\phi$ mesons.
\end{itemize}

\baeeq
\label{e101ap03}
\Pi^{\Sigma^{\ast 0} \rar \Sigma^0 \omega} \es {1 \over \sqrt{2}}
[\Pi_1(u,d,s) + \Pi_1(d,u,s) ]~, \nnb \\
\Pi^{\Sigma^{\ast +} \rar \Sigma^+ \omega} \es - \sqrt{2} \Pi_1(u,u,s )~, \nnb \\
\Pi^{\Sigma^{\ast -} \rar \Sigma^- \omega} \es \sqrt{2} \Pi_1(d,d,s )~, \nnb \\
\Pi^{\Delta^+ \rar p \omega} \es - \sqrt{2}
[\Pi_1(d,u,u) - \Pi_1(u,u,d) ]~, \nnb \\
\Pi^{\Delta^0 \rar n \omega} \es \sqrt{2}
[ \Pi_1(u,d,d) - \Pi_1(d,d,u) ]~, \nnb \\
\Pi^{\Xi^{\ast 0} \rar \Xi^0 \omega} \es - \sqrt{2} \Pi_1(u,s,s)~, \nnb \\
\Pi^{\Xi^{\ast -} \rar \Xi^- \omega} \es \sqrt{2} \Pi_1(d,s,s)~, \nnb \\
\Pi^{\Sigma^{\ast 0} \rar \Lambda \omega} \es - {1 \over \sqrt{6}}
[2 \Pi_1(u,s,d) -2 \Pi_1(d,s,u) - \Pi_1(d,u,s) +
\Pi_1(u,d,s) ]~, \nnb \\
\Pi^{\Sigma^{\ast 0} \rar \Sigma^0 \phi} \es 
- [\Pi_1(s,d,u) + \Pi_1(s,u,d)]~, \nnb \\
\Pi^{\Sigma^{\ast +} \rar \Sigma^+ \phi} \es  2 \Pi_1(s,u,u)~, \nnb \\
\Pi^{\Sigma^{\ast -} \rar \Sigma^- \phi} \es -2 \Pi_1(s,d,d)~, \nnb \\
\Pi^{\Delta^+ \rar p \phi} \es \Pi^{\Delta^0 \rar n \phi} = 0~, \nnb \\
\Pi^{\Xi^{\ast 0} \rar \Xi^0 \phi} \es 2 \Pi_1(s,s,u )~, \nnb \\
\Pi^{\Xi^{\ast -} \rar \Xi^- \phi} \es - 2 \Pi_1(s,s,d )~, \nnb \\
\Pi^{\Sigma^{\ast 0} \rar \Lambda \phi} \es - {1\over \sqrt{3}}
[ \Pi_1(s,u,d) - \Pi_1(s,d,u) ]~.
\eaeeq

Obviously, in the $SU(2)$ symmetry limit, $\Pi^{\Delta^+ \rar \Lambda
K^{\ast +}}$, $\Pi^{\Delta^0 \rar \Lambda \bar{K}^{\ast 0}}$, $\Pi^{\Delta^+
\rar p \omega}$, $\Pi^{\Delta^0 \rar n \omega}$, $\Pi^{\Sigma^{\ast 0}
\rar \Lambda \omega}$ and $\Pi^{\Sigma^{\ast 0} \rar \Lambda \phi}$ are all
equal to zero. Note that, in calculating $\Pi^{D \rar O V}$ using Eqs.
(\ref{e101ap01})--(\ref{e101ap03}), one should use the wave functions of the
corresponding vector meson in $\Pi_1$. 

\eAPP

\newpage
\bAPP{B}{}

In this appendix we present the DA's of the vector mesons
appearing in the matrix elements $\lla V(q) \vel \bar{q}(x) \Gamma q(0) \ver
0 \rra$ and $\lla V(q) \vel \bar{q}(x) G_{\mu\nu} q(0) \ver 0 \rra$,
up to twist--4 accuracy \cite{R10110,R10111,R10112}:

\baeeq
\label{e101apb01}
\lla V(q,\lambda) \vel \bar{q}_1(x) \gamma_\mu q_2(0) \ver 0 \rra \es
f_V m_V \Bigg\{ {\varepsilon^\lambda \mcdot x \over q\mcdot x} q_\mu 
\int_0^1 du e^{i \bar{u} q\mcdot x} \Bigg[ \phi_\parallel (u) + {m_V^2 x^2 \over 16}
A_\parallel (u) \Bigg] \nnb \\
\ar \Bigg( \varepsilon_\mu^\lambda - q_\mu {\varepsilon^\lambda \mcdot x \over q\mcdot x} \Bigg) \int_0^1
du e^{i \bar{u} q\mcdot x} g_\perp^v (u) \nnb \\
\ek {1\over 2} x_\mu { \varepsilon^\lambda \mcdot x
\over (q\mcdot x)^2} m_V^2 \int_0^1 du e^{i \bar{u} q\mcdot x} \Big[ g_3 (u) +
\phi_\parallel (u) - 2 g_\perp^v (u) \Big] \Bigg\}~, \nnb \\ \nnb \\
\lla V(q,\lambda) \vel \bar{q}_1(x) \gamma_\mu \gamma_5 q_2(0) \ver 0 \rra
\es - {1 \over 4} \epsilon_\mu^{\nu\alpha\beta} \varepsilon^\lambda q_\alpha
x_\beta f_V m_V \int_0^1 du e^{i \bar{u} q\mcdot x} g_\perp^a (u)~, 
\nnb \\ \nnb \\
\lla V(q,\lambda) \vel \bar{q}_1(x) \sigma_{\mu\nu} q_2(0) \ver 0 \rra \es  
- i f_V^T \Bigg\{ (\varepsilon_\mu^\lambda q_\nu - \varepsilon_\nu^\lambda
q_\mu ) \int_0^1 du e^{i \bar{u} q\mcdot x}\Bigg[\phi_\perp (u) + {m_V^2
x^2 \over 16} A_\perp (u) \Bigg] \nnb \\
\ar { \varepsilon^\lambda \mcdot x \over (q\mcdot x)^2} (q_\mu x_\nu - q_\nu
x_\mu) \int_0^1 du e^{i \bar{u} q\mcdot x} \Bigg[h_\parallel^t - {1\over 2}
\phi_\perp - {1\over 2} h_3 (u) \Bigg] \nnb \\
\ar {1\over 2} (\varepsilon_\mu^\lambda x_\nu - \varepsilon_\nu^\lambda
x_\mu) {m_V^2 \over q \mcdot x} \int_0^1 du e^{i \bar{u} q\mcdot x}
\Big[h_3(u) - \phi_\perp (u) \Big] \Bigg\}~, \nnb \\ \nnb \\
\lla V(q,\lambda) \vel \bar{q}_1(x) \sigma_{\alpha\beta} g G_{\mu\nu}(u x) q_2(0) \ver 0 \rra \es  
f_V^T m_V^2 { \varepsilon^\lambda \mcdot x \over 2 q\mcdot x} \Big[q_\alpha q_\mu
g_{\beta\nu}^\perp - q_\beta q_\mu g_{\alpha\nu}^\perp - q_\alpha q_\nu g_{\beta\mu}^\perp
+ q_\beta q_\nu g_{\alpha\mu}^\perp \Big] \nnb \\
\cp \int {\cal D}\alpha_i e^{i(\alpha_{\bar{q}}
+ u \alpha_g) q\mcdot x} {\cal T}(\alpha_i) \nnb \\ 
\ar f_V^T m_V^2 \Big[q_\alpha \varepsilon_\mu^\lambda g_{\beta\nu}^\perp - q_\beta
\varepsilon_\mu^\lambda g_{\alpha\nu}^\perp - q_\alpha
\varepsilon_\nu^\lambda g_{\beta\mu}^\perp
+ q_\beta \varepsilon_\nu^\lambda g_{\alpha\mu}^\perp \Big] \nnb \\
\cp \int {\cal D}\alpha_i e^{i(\alpha_{\bar{q}}
+ u \alpha_g) q\mcdot x} {\cal T}_1^{(4)}(\alpha_i) \nnb \\
\ar f_V^T m_V^2 \Big[q_\mu \varepsilon_\alpha^\lambda g_{\beta\nu}^\perp -
q_\mu \varepsilon_\beta^\lambda g_{\alpha\nu}^\perp - q_\nu
\varepsilon_\alpha^\lambda g_{\beta\mu}^\perp
+ q_\nu \varepsilon_\beta^\lambda g_{\alpha\mu}^\perp \Big] \nnb \\
\cp \int {\cal D}\alpha_i e^{i(\alpha_{\bar{q}}
+ u \alpha_g) q\mcdot x} {\cal T}_2^{(4)}(\alpha_i) \nnb \\
\ar {f_V^T m_V^2 \over q \mcdot x} \Big[q_\alpha q_\mu \varepsilon_\beta^\lambda
x_\nu - q_\beta q_\mu \varepsilon_\alpha^\lambda x_\nu -
q_\alpha q_\nu \varepsilon_\beta^\lambda x_\mu +
q_\beta q_\nu \varepsilon_\alpha^\lambda x_\mu \nnb \\
\cp \int {\cal D}\alpha_i e^{i(\alpha_{\bar{q}}
+ u \alpha_g) q\mcdot x} {\cal T}_3^{(4)}(\alpha_i) \nnb \\
\ar {f_V^T m_V^2 \over q \mcdot x} \Big[q_\alpha q_\mu \varepsilon_\nu^\lambda
x_\beta - q_\beta q_\mu \varepsilon_\nu^\lambda x_\alpha -
q_\alpha q_\nu \varepsilon_\mu^\lambda x_\beta +
q_\beta q_\nu \varepsilon_\mu^\lambda x_\alpha \nnb \\
\cp \int {\cal D}\alpha_i e^{i(\alpha_{\bar{q}}
+ u \alpha_g) q\mcdot x} {\cal T}_4^{(4)}(\alpha_i)~, \nnb \\ \nnb \\
\lla V(q,\lambda) \vel \bar{q}_1(x) g_s G_{\mu\nu} (ux) q_2(0) \ver 0 \rra \es
-i f_V^T m_V (\varepsilon_\mu^\lambda q_\nu - \varepsilon_\nu^\lambda q_\mu)
\int {\cal D}\alpha_i e^{i(\alpha_{\bar{q}}
+ u \alpha_g) q\mcdot x} {\cal S}(\alpha_i)~, \nnb \\ \nnb \\
\lla V(q,\lambda) \vel \bar{q}_1(x) g_s \widetilde{G}_{\mu\nu} (ux) 
\gamma_5 q_2(0) \ver 0 \rra \es
-i f_V^T m_V (\varepsilon_\mu^\lambda q_\nu - \varepsilon_\nu^\lambda q_\mu)
\int {\cal D}\alpha_i e^{i(\alpha_{\bar{q}}
+ u \alpha_g) q\mcdot x} \widetilde{{\cal S}}(\alpha_i)~, \nnb \\ \nnb \\
\lla V(q,\lambda) \vel \bar{q}_1(x) g_s \widetilde{G}_{\mu\nu} (ux) 
\gamma_\alpha \gamma_5 q_2(0) \ver 0 \rra \es
f_V m_V q_\alpha (\varepsilon_\mu^\lambda q_\nu - \varepsilon_\nu^\lambda q_\mu)
\int {\cal D}\alpha_i e^{i(\alpha_{\bar{q}}
+ u \alpha_g) q\mcdot x} {\cal A}(\alpha_i)~, \nnb \\ \nnb \\
\lla V(q,\lambda) \vel \bar{q}_1(x) g_s G_{\mu\nu} (ux) i \gamma_\alpha 
q_2(0) \ver 0 \rra \es
f_V m_V q_\alpha (\varepsilon_\mu^\lambda q_\nu - \varepsilon_\nu^\lambda q_\mu)
\int {\cal D}\alpha_i e^{i(\alpha_{\bar{q}}
+ u \alpha_g) q\mcdot x} {\cal V}(\alpha_i)~,
\eaeeq
where $\widetilde{G}_{\mu\nu} = (1/2) \epsilon_{\mu\nu\alpha\beta}
G^{\alpha\beta}$ is the dual gluon field strength tensor, and $\int {\cal D}
\alpha_i = \int d\alpha_q d\alpha_{\bar{q}} d\alpha_g \delta (1 - \alpha_q -
\alpha_{\bar{q}} - \alpha_g)$. 

\eAPP

\newpage

\bAPP{C}{}

\baeeq
\label{e101}
\Pi_1^{(1)} \es
{M^4\over 192\sqrt{3} \pi^2}  \Big\{    
-12 f_V^\perp [m_s (1+\beta) +2 m_d \beta] \phi_V^\perp(u_0) \nnb \\
\ar f_V^\parallel m_V (1-\beta) [8 i_3({\cal A},1-v) + 12
\psi_{3;V}^\perp(u_0) - \psi_{3;V}^{\perp\prime}(u_0)
- 2 \Bbb{B}(u_0)] \Big\} \nnb \\
\ar \Bigg(\gamma_E - \ln {M^2\over \Lambda^2}\Bigg) 
\Bigg\{ {f_V^\perp m_V^2 M^2\over 24 \sqrt{3} \pi^2} 
\Big( [m_s (1+3\beta)-2 m_d \beta] i_2({\cal S},1) \nnb \\
\ar [m_s (3+\beta)-2 m_d] i_2(\widetilde{\cal S},1)
+ 2 [m_d (1-\beta) + m_s (1+2 \beta)] i_2({\cal T}_1,1) \nnb \\
\ar 2 [m_d (1-\beta) - m_s (2+\beta)] i_2({\cal T}_2,1)
+ [2 m_d \beta + m_s (1+\beta)] i_2({\cal T}_3,1) \nnb \\
\ar [2 m_d \beta - m_s (1+3 \beta)] i_2({\cal T}_4,1) \Big) \nnb \\
\ek {f_V^\perp m_V^4\over 24 \sqrt{3} \pi^2}  \Big(
[2 m_d \beta + m_s (1+\beta)] i_1({\cal T}_3,1)
+ [2 m_d \beta - m_s (1+3 \beta)] i_1({\cal T}_4,1) \Big) \nnb \\
\ek {f_V^\perp \GG\over 384 \sqrt{3} \pi^2 M^2}  \Big(m_V^2 [2 m_d - m_s (3+\beta)]
\widetilde{i}_4(\Bbb{C}_T)
- 2 M^2 [2 m_d \beta + m_s (1+\beta)] \phi_V^\perp(u_0) \Big)\Bigg\} \nnb \\
\ar {f_V^\parallel m_0^2 m_V \GG\over 144 \sqrt{3} M^6}  
\beta [m_d \sp + m_s \dd] \psi_{3;V}^\perp(u_0) \nnb \\
% qsq ??
%
\ek {m_V\over 1728 \sqrt{3} \pi^2 M^4} \Big\{
3 f_V^\perp m_V^3 \GG [2 m_d \beta - m_s (1+3 \beta)]
i_1({\cal T}_4,1) \nnb \\
\ar 4 \pi^2  f_V^\parallel \Big( \sp [m_0^2 m_s m_V^2 (1 - \beta) -
6 \GG m_d \beta]
+ \dd [m_0^2 m_d m_V^2 (1 - \beta) \nnb \\
\ek 6 \GG m_s \beta]
+ m_0^2 m_V^2 (1-\beta) [m_d \dd + m_s \sp] \Big)
\psi_{3;V}^\perp(u_0) \Big\}\nnb \\
% 
% BU KISIM YUKARIYA YAZILDI(1)
%\ar {i\over 576 \sqrt{3} \pi^2 M^4} \GG [2 m_d \beta - m_s (1+3 \beta)]
%f_V^\perp m_V^4 i_1({\cal T}_4,1) \nnb \\
%\ar {i\over 432 \sqrt{3} M^4} \Big\{ \sp [m_0^2 m_s m_V^2 (1 - \beta) - 
%6 \GG m_d \beta] +
%\dd [m_0^2 m_d m_V^2 (1 - \beta) \nnb \\
%\ek 6 \GG m_s \beta]
%+ m_0^2 (m_d \dd + m_s \sp) (1-\beta) m_V^2 \Big\} f_V^\parallel m_V
%\psi_{3;V}^\perp(u_0) \nnb \\
%
\ar {f_V^\perp m_V^2 \GG\over 576 \sqrt{3} \pi^2 M^2}  \Big\{
2 \beta (2 m_d - m_s) i_2({\cal S},1) +
2 [m_d (1+\beta) - m_s] i_2(\widetilde{\cal S},1- 2 v) \nnb \\
\ek 2 \beta m_d i_2({\cal T}_1,1+ 2 v) -
2 m_s (1+\beta) i_2({\cal T}_1,v) +
 m_s (1+3 \beta) i_2({\cal T}_1,1) \nnb \\
\ar [m_s (3+\beta) - 2 m_d] i_2({\cal T}_2,1) +
4 [m_d (1+\beta) - m_s] i_2({\cal T}_2,v) \nnb \\
\ar [m_s (1+\beta) +2 \beta m_d] i_2({\cal T}_3,1- 2 v)
+ 4 \beta m_d i_2({\cal T}_4,v) \nnb \\
\ek 2 m_s [ i_2({\cal T}_4,1-v) + \beta i_2({\cal T}_4,2-v)] \Big\} \nnb \\
%
%  BU KISIM YUKARIYA YAZILDI(2)
%\ek {i\over 288 \sqrt{3} \pi^2 M^2} \GG \beta (2 m_d - m_s) f_V^\perp
%m_V^2 i_2({\cal S},1) \nnb \\
%\ek {i \over 288 \sqrt{3} \pi^2 M^2} \GG [m_d (1+\beta) - m_s] f_V^\perp
%m_V^2 i_2(\widetilde{\cal S},1- 2 v) \nnb \\
%\ar {i\over 576 \sqrt{3} \pi^2 M^2} \GG [2 m_d \beta i_2({\cal T}_1,1+ 2 v) +
%2 m_s (1+\beta) i_2({\cal T}_1,v) \nnb \\
%\ek  m_s (1+3 \beta) i_2({\cal T}_1,1)] f_V^\perp
%m_V^2 \nnb \\
%\ek {i\over 576 \sqrt{3} \pi^2 M^2} \GG \{ [m_s (3+\beta) - 2 m_d] 
%i_2({\cal T}_2,1) \nnb \\
%\ar 4 [m_d (1+\beta) - m_s] i_2({\cal T}_2,v) \} f_V^\perp m_V^2 \nnb \\
%\ek {i\over 576 \sqrt{3} \pi^2 M^2} \GG [m_s (1+\beta) +2 \beta m_d]
%f_V^\perp m_V^2 i_2({\cal T}_3,1- 2 v) \nnb \\
%\ek {i\over 288 \sqrt{3} \pi^2 M^2} \GG \{ 2 m_d \beta i_2({\cal T}_4,v) -
%m_s [ i_2({\cal T}_4,1-v) + \beta i_2({\cal T}_4,2-v)] \} f_V^\perp m_V^2 \nnb \\
%
%
\ek {1\over 432 \sqrt{3} M^2} \Big\{
24 f_V^\perp m_V^4 [2 \beta \dd - (1+3 \beta) \sp] i_1({\cal T}_4,1) \nnb \\
\ar 6 f_V^\parallel m_V^3 (1-\beta)
\Big( [m_d \dd - m_s \sp] i_2({\cal V},5-4 v) - 
[m_d \dd + m_s \sp] i_2({\cal A},5-6 v) \Big) \nnb \\
\ek f_V^\parallel m_0^2 m_V (1-\beta) [m_d \dd + m_s \sp] [ 2 \Bbb{B}(u_0) +
\psi_{3;V}^{\perp\prime}(u_0)] \nnb \\ 
\ar 2 f_V^\parallel m_V 
\Big( 2 m_0^2 (1+\beta) [m_d \sp + m_s \dd] - 9 m_V^2 (1-\beta) 
(m_d \dd + m_s \sp) \Big)
\psi_{3;V}^\perp(u_0) \Big\}\nnb \\
% 
% BU KISIM YUKARIYA YAZILDI(3)
%\ar {i\over 18 \sqrt{3} M^2} [2 \beta \dd - (1+3 \beta) \sp] f_V^\perp
%m_V^4 i_1({\cal T}_4,1) \nnb \\
% qsq ??
%\ek {i\over 72 \sqrt{3} M^2} (1-\beta) (m_d \dd + m_s \sp) f_V^\parallel
%m_V^3 i_2({\cal A},(5-6 v)) \nnb \\
%\ar {i\over 72 \sqrt{3} M^2} (1-\beta) (m_d \dd - m_s \sp) f_V^\parallel
%m_V^3 i_2({\cal V},5-4 v) \nnb \\
%\ek {i\over 432 \sqrt{3} M^2} m_0^2 (m_d \dd + m_s \sp) (1-\beta)
%f_V^\parallel m_V [ 2 \Bbb{B}(u_0) + \psi_{3;V}^{\perp\prime}(u_0)] \nnb \\ 
%\ar {i\over 216 \sqrt{3} M^2} [ 2 m_0^2 (m_d \sp + m_s \dd) (1+\beta)
%- 9 (1-\beta) (m_d \dd + m_s \sp) m_V^2 ] 
%f_V^\parallel m_V \psi_{3;V}^\perp(u_0) \nnb \\
%
\ek {f_V^\perp m_V^2\over 9216 \sqrt{3} \pi^2 M^2}  \Big\{
16 \Big( \GG [2 m_d - m_s (3+ \beta)] \nnb \\
\ek 2 \pi^2 m_0^2 [ \dd (5-\beta) - 2 \sp (5+2 \beta)] \Big) 
\widetilde{i}_4(\Bbb{C}_T) + 9 \GG [2 \beta m_d + m_s (1+\beta)]
\Bbb{A}_T(u_0) \Big\}\nnb \\
%
%  BU KISIM YUKARIYA YAZILDI(4)
%\ar {i\over 576 \sqrt{3} \pi^2 M^2} \GG [2 m_d - m_s (3+ \beta)]
%f_V^\perp m_V^2 \widetilde{i}_4(\Bbb{C}_T) \nnb \\
%\ek {i\over 216 \sqrt{3} M^2} [ \dd (5-\beta) - 2 \sp (5+2 \beta)]
%f_V^\perp m_0^2 m_V^2 \widetilde{i}_4(\Bbb{C}_T) \nnb \\
%\ar {i\over 768 \sqrt{3} \pi^2 M^2} \GG [2 \beta m_d + m_s (1+\beta)]
%f_V^\perp m_V^2 \Bbb{A}_T(u_0) \nnb \\
%
%
%*********************************************************************
%
\ek {M^2\over 96 \sqrt{3} \pi^2} \Big\{
f_V^\perp m_V^2 \Big(
- 3 [2 \beta m_d + m_s (1+\beta)] \Bbb{A}_T(u_0) +
4 [6 \beta m_d - m_s (1+5 \beta)] i_2({\cal S},1) \nnb \\
\ar 4 [2 m_d \beta + m_s (1+\beta)]
i_2(\widetilde{\cal S},1) - 16 [m_d (1+\beta) - m_s] i_2(\widetilde{\cal
S},v) \nnb \\
\ar 4 [2 m_d + m_s (3+7 \beta)]
i_2({\cal T}_1,1) - 16 m_d \beta i_2({\cal T}_1,1+v)
- 8 m_s (1+\beta) i_2({\cal T}_1,v) \nnb \\
\ek 4 [2 m_d \beta + m_s (1+\beta)]
i_2({\cal T}_2,1) - 16 [m_d (1+\beta) - m_s] i_2({\cal T}_2,v) \nnb \\
\ar 8 [2 m_d \beta + m_s (1+\beta)]
i_2({\cal T}_3,1-v)
+ 4 [2 m_d \beta - m_s (3+7\beta)] i_2({\cal T}_4,1) \nnb \\
\ar 8 [2 m_d \beta + m_s (1+\beta)] i_2({\cal T}_4,v)
- 3 [2 m_d - m_s (3+\beta)]  \widetilde{i}_4(\Bbb{C}_T) \Big) \nnb \\
\ek 16 \pi^2 f_V^\perp [2 \beta \dd + \sp (1+\beta)] \phi_V^\perp(u_0)
- f_V^\parallel m_V^3 (1-\beta) [ 2 i_2({\cal A},5-6 v) +
\psi_{3;V}^\perp(u_0) ] \Big\} \nnb \\
%---------------------------------------------------------------------
%
% BU KISIM YUKARIYA YAZILDI(5)
%\ek {i\over 32 \sqrt{3} \pi^2} M^2 [2 \beta m_d + m_s (1+\beta)] f_V^\perp
%m_V^2 \Bbb{A}_T(u_0) \nnb \\
%\ar {i\over 24 \sqrt{3} \pi^2} M^2 [6 \beta m_d - m_s (1+5 \beta)]
%f_V^\perp m_V^2 i_2({\cal S},1) \nnb \\
%\ar {i\over 24 \sqrt{3} \pi^2} M^2 \{ [2 m_d \beta + m_s (1+\beta)]
%i_2(\widetilde{\cal S},1) - 4 [m_d (1+\beta) - m_s] i_2(\widetilde{\cal S},v) \} f_V^\perp
%m_V^2 \nnb \\
%\ar {i\over 24 \sqrt{3} \pi^2} M^2 \{ [2 m_d + m_s (3+7 \beta)]      
%i_2({\cal T}_1,1) - 4 m_d \beta i_2({\cal T}_1,1+v) \nnb \\
%\ek 2 m_s (1+\beta) i_2({\cal T}_1,v) \}
%f_V^\perp m_V^2 \nnb \\
%\ek {i\over 24 \sqrt{3} \pi^2} M^2 \{ [2 m_d \beta + m_s (1+\beta)]
%i_2({\cal T}_2,1) - 4 [m_d (1+\beta) - m_s] i_2({\cal T}_2,v) \} f_V^\perp
%m_V^2 \nnb \\
%\ar {i\over 12 \sqrt{3} \pi^2} M^2 \{ [2 m_d \beta + m_s (1+\beta)]
%i_2({\cal T}_3,1-v)  f_V^\perp m_V^2 \nnb \\
%\ar {i\over 24 \sqrt{3} \pi^2} M^2 \{ [2 m_d \beta - m_s (3+7\beta)] i_2({\cal T}_4,1)
%+ 2 [2 m_d \beta + m_s (1+\beta)] i_2({\cal T}_4,v) \} f_V^\perp m_V^2 \nnb \\
%
%\ek {i\over 48 \sqrt{3} \pi^2} M^2 f_V^\parallel m_V^3
%(1-\beta) i_2({\cal A},5-6 v) \nnb \\
%
%\ek {i\over 32 \sqrt{3} \pi^2} M^2 [2 m_d - m_s (3+\beta)] f_V^\perp
%m_V^2 \widetilde{i}_4(\Bbb{C}_T) \nnb \\
%\ek {i\over 6 \sqrt{3}} M^2 [2 \beta \dd + \sp (1+\beta)] f_V^\perp     
%\phi_V^\perp(u_0) \nnb \\
%\ek {i\over 96 \sqrt{3} \pi^2} M^2 (1-\beta) f_V^\parallel m_V^3
%\psi_{3;V}^\perp(u_0) \nnb \\
%
%*********************************************************************
%
%
\ek {m_V f_V^\parallel\over 144 \sqrt{3}}  \Big\{
2 (1-\beta) [m_d \dd + m_s \sp] [3\Bbb{B}(u_0) - 2i_3({\cal A},1-v) +
\psi_{3;V}^{\perp\prime}(u_0)] \nnb \\
\ar 4 (1-\beta) [m_d \dd - m_s \sp]  i_3({\cal V},1-v) \nnb \\
\ek 12 \Big( [4 \beta m_d
+(1-\beta) m_s] \sp + [(1-\beta) m_d + 4 \beta m_s] \dd \Big)
\psi_{3;V}^\perp(u_0) \Big\}\nnb \\
%
% BU KISIM YUKARIYA YAZILDI(6)
%\ar {i\over 24 \sqrt{3}} m_V f_V^\parallel (1-\beta) 
%(m_d \dd + m_s \sp) \Bbb{B}(u_0) \nnb \\
%\ek {i\over 36 \sqrt{3}} m_V f_V^\parallel (1-\beta) 
%(m_d \dd + m_s \sp) i_3({\cal A},1-v) \nnb \\
%\ar {i\over 36 \sqrt{3}} m_V f_V^\parallel (1-\beta) 
%(m_d \dd - m_s \sp) i_3({\cal V},1-v) \nnb \\
%\ek {i\over 12 \sqrt{3}} f_V^\parallel m_V \{ [4 \beta m_d
%+(1-\beta) m_s] \sp + [(1-\beta) m_d + 4 \beta m_s] \dd \} 
%\psi_{3;V}^\perp(u_0) \nnb \\
%\ar {i\over 48 \sqrt{3}} f_V^\parallel m_V (1-\beta) (m_d \dd + m_s
%\sp) \psi_{3;V}^{\perp\prime}(u_0) \nnb \\
%*********************************************************************
\ek {f_V^\perp m_V^2\over 72 \sqrt{3}}  \Big\{
[2 \beta \dd + (1+\beta) \sp] [ 3\Bbb{A}_T(u_0) + 8 i_2({\cal T}_1,v) 
- 4 i_2({\cal T}_3,1-2 v) - 8 i_2({\cal T}_4,v)] \nnb \\
\ek 8 \beta [2 \dd - \sp] i_2({\cal S},1)
- 8 [(1+\beta) \dd - \sp] i_2(\widetilde{\cal S},1-2 v) \nnb \\
\ar 4 [2 \beta \dd - (1+3 \beta) \sp] i_2({\cal T}_1,1) +
4 [2 \dd - (3+\beta) \sp] i_2({\cal T}_2,1) \nnb \\
\ek 16 [(1+\beta) \dd -\sp] i_2({\cal T}_2,v)
+ 8 (1+2 \beta) \sp i_2({\cal T}_4,1) 
+ 6 [2 \dd -(3+\beta) \sp] \widetilde{i}_4(\Bbb{C}_T) \nnb \\
\ek {3 m_V^2\over \pi^2} \Big( [2 \beta m_d+(1+\beta) m_s] i_1({\cal T}_3,1) + 
2 [2 \beta m_d -(1+3 \beta) m_s] i_1({\cal T}_4,1) \Big) \Big\} \nnb \\
% 
%---------------------------------------------------------------------
%
%BU KISIM YUKARIYA YAZILDI(7)
%\ar {i\over 24 \sqrt{3}} f_V^\perp m_V^2 [2 \beta \dd + (1+\beta) \sp]
%\Bbb{A}_T(u_0) \nnb \\
%\ek {i\over 9 \sqrt{3}} f_V^\perp m_V^2 \beta (2 \dd - \sp)
%i_2({\cal S},1) \nnb \\
%\ek {i\over 9 \sqrt{3}} f_V^\perp m_V^2 [(1+\beta) \dd - \sp]
%i_2(\widetilde{\cal S},1-2 v) \nnb \\
%\ar {i\over 18 \sqrt{3}} f_V^\perp m_V^2 \{ [2 \beta \dd - (1+3 \beta) \sp]
%i_2({\cal T}_1,1) + 2 [ 2 \beta \dd + (1+\beta) \sp ] i_2({\cal T}_1,v) \}
%\nnb \\
%\ar {i\over 18 \sqrt{3}} f_V^\perp m_V^2 \{ [2 \dd - (3+\beta) \sp]
%i_2({\cal T}_2,1) - 4 [(1+\beta) \dd -\sp] i_2({\cal T}_2,v) \} \nnb \\
%\ek {i\over 18 \sqrt{3}} f_V^\perp m_V^2 [2 \beta \dd (1+\beta) \sp]
%i_2({\cal T}_3,1-2 v) \nnb \\
%\ar {i\over 9 \sqrt{3}} f_V^\perp m_V^2 \{
%(1+2 \beta) \sp i_2({\cal T}_4,1) - [2 \beta \dd + (1+\beta) \sp]
%i_2({\cal T}_4,v) \} \nnb \\
%\ek {i\over 24 \sqrt{3} \pi^2} f_V^\perp m_V^4 [2 \beta m_d
%+(1+\beta) m_s] i_1({\cal T}_3,1) \nnb \\
%\ek {i\over 12 \sqrt{3} \pi^2} f_V^\perp m_V^4 [2 \beta m_d -(1+3
%\beta) m_s] i_1({\cal T}_4,1) \nnb \\
%\ar {i\over 12 \sqrt{3}} f_V^\perp m_V^2 [2 \dd -(3+\beta) \sp]
%\widetilde{i}_4(\Bbb{C}_T) \nnb \\
%
%*********************************************************************
%
\ar {f_V^\perp\over 864 \sqrt{3} \pi^2}  \Big\{
3 \GG [2 \beta m_d + (1+\beta) m_s]   
-  40 \pi^2 m_0^2
[2 \beta \dd + (1+\beta) \sp] \Big\} \phi_V^\perp(u_0)~, \nnb \\ 
%
%*********************************************************************
%
%%BU KISIM YUKARIYA YAZILDI(8) 
%\ek {i\over 288 \sqrt{3} \pi^2} f_V^\perp \GG [2 \beta m_d + (1+\beta) m_s]
%\phi_V^\perp(u_0) \nnb \\
%\ar  {5 i\over 108 \sqrt{3}} m_0^2 f_V^\perp
%[2 \beta \dd + (1+\beta) \sp] \phi_V^\perp(u_0)~, \nnb \\
\eaeeq

\baeeq
\label{e102}
\Pi_1^{(2)} \es
\Bigg(\gamma_E - \ln {M^2\over \Lambda^2}\Bigg) \Bigg\{
-{f_V^\perp m_V^2\over 12 \sqrt{3} \pi^2}  (1-\beta)
[(2 m_d -m_s) i_1({\cal T},1) - 2 m_di_1({\cal T}_3,1) \nnb \\
\ek 2 (m_d - m_s) i_1({\cal T}_4,1) ]
- {f_V^\perp m_V^2 \GG \over 96 \sqrt{3} \pi^2 M^4} [2 m_d - m_s (3 + \beta)]
\widetilde{\widetilde{i}}_4(\Bbb{B}_T)
\Bigg\} \nnb \\
%%BU KISIM YUKARIYA YAZILDI(9)
%
%{i\over 12 \sqrt{3} \pi^2} f_V^\perp m_V^2
%(1-\beta) (2 m_d -m_s) i_1({\cal T},1) \nnb \\
%\ek {i\over 6 \sqrt{3} \pi^2} m_d (1-\beta) f_V^\perp 
%m_V^2 i_1({\cal T}_3,1) \nnb \\
%\ek {i\over 6 \sqrt{3} \pi^2} (m_d - m_s) (1-\beta) 
%f_V^\perp m_V^2 i_1({\cal T}_4,1) \nnb \\
%\ar {i\over 96 \sqrt{3} \pi^2 M^4} \GG [2 m_d - m_s (3 + \beta)]
%f_V^\perp m_V^2 \widetilde{\widetilde{i}}_4(\Bbb{B}_T)
%\Bigg\} \nnb \\
%
\ar {f_V^\parallel m_V^3 m_0^2\over 18\sqrt{3} M^6}  
\Big\{[m_d \sp (3+\beta) - m_s \dd (1+\beta)]
i_0(\Psi,1) \nnb \\
\ar [m_d \sp (1+3\beta) - m_s \dd (1+\beta)] 
i_0(\widetilde{\Psi},1) \Big\} \nnb \\
\ek {f_V^\perp m_V^2 \GG\over 288\sqrt{3} \pi^2 M^4}  \Big\{
(m_d - 2 m_s) (1+\beta) [i_1({\cal T},1)  - 2 i_1({\cal T}_4,1) ] \nnb \\
\ek [2 m_d - m_s (3 + \beta)] (3 + \beta) [i_1({\cal T},v) - 
2 i_1({\cal T}_4,v)] \Big\} \nnb \\
%
%
%BU KISIM YUKARIYA YAZILDI(10)
%\ar {i\over 288\sqrt{3} \pi^2 M^4} \GG (m_d - 2 m_s) (1+\beta) m_V^2
%f_V^\perp i_1({\cal T},1) \nnb \\
%\ek {i\over 288\sqrt{3} \pi^2 M^4} \GG [2 m_d - m_s (3 + \beta)] m_V^2
%f_V^\perp i_1({\cal T},v) \nnb \\
%\ek {i\over 144\sqrt{3} \pi^2 M^4} \GG (m_d - 2 m_s) (1+\beta) m_V^2
%f_V^\perp i_1({\cal T}_4,1) \nnb \\
%\ar {i\over 144\sqrt{3} \pi^2 M^4} \GG [2 m_d - m_s (3 + \beta)] m_V^2
%f_V^\perp i_1({\cal T}_4,v) \nnb \\
%
\ar {2 f_V^\parallel m_V^3 \over 9\sqrt{3} M^4} [m_d \sp (3+\beta) - m_s \dd (1+\beta)] 
i_0(\Psi,1) \nnb \\
\ar {f_V^\parallel m_V^3\over 144 \sqrt{3} M^4}  (1-\beta)
\Big\{ 4 [m_d \dd - m_s \sp] [i_1(\Phi,1) + i_1(\widetilde{\Psi},1)] \nnb \\
\ek [m_d \dd + m_s \sp] [4 i_1(\widetilde{\Phi},1) 
+ 4 i_1(\Psi,1-2 v) + 3 \widetilde{i}_4({\Bbb A})] \nnb \\
\ar 8 [2 m_d \dd - m_s \sp] i_1(\Phi,v)
+ 8 [m_d \dd + 2 m_s \sp] i_1(\widetilde{\Phi},v) \Big\}\nnb \\
\ar {f_V^\parallel m_V m_0^2\over 216 \sqrt{3} M^4}  
[m_d \dd + m_s \sp] (1-\beta) [
2 \widetilde{i}_4(\Bbb{B}) - 2 \widetilde{i}_4(\phi_V^\parallel) +
\psi_{3;V}^\perp(u_0) ] \nnb \\
\ar {f_V^\perp m_V^2\over 432 \sqrt{3} \pi^2 M^4}  \Big\{
\GG [6 m_d - 3 m_s (3+\beta)] -
8 \pi^2 m_0^2 [ \dd (5-\beta) - 2 \sp (5 + 2 \beta)] \Big\}
\widetilde{\widetilde{i}}_4(\Bbb{B}_T) \nnb \\
%
%%BU KISIM YUKARIYA YAZILDI(11)
%\ek {2 i\over 9\sqrt{3} M^4} [m_d \sp (3+\beta) - m_s \dd (1+\beta)]
%f_V^\parallel m_V^3 i_0(\Psi,1) \nnb \\
%\ek {i\over 36 \sqrt{3} M^4} (m_d \dd - m_s \sp) (1-\beta) m_V^3
%f_V^\parallel i_1(\Phi,1) \nnb \\
%\ek {i\over 18 \sqrt{3} M^4} (2 m_d \dd - m_s \sp) (1-\beta) m_V^3
%f_V^\parallel i_1(\Phi,v) \nnb \\
%\ar {i\over 36 \sqrt{3} M^4} (m_d \dd + m_s \sp) (1-\beta) m_V^3
%f_V^\parallel i_1(\widetilde{\Phi},1) \nnb \\ 
%\ek {i\over 18 \sqrt{3} M^4} (m_d \dd + 2 m_s \sp) (1-\beta) m_V^3
%f_V^\parallel i_1(\widetilde{\Phi},v) \nnb \\
%\ar {i\over 36 \sqrt{3} M^4} (m_d \dd + m_s \sp) (1-\beta) m_V^3
%f_V^\parallel i_1(\Psi,1-2 v) \nnb \\
%\ek {i\over 36 \sqrt{3} M^4} (m_d \dd - m_s \sp) (1-\beta) m_V^3
%f_V^\parallel i_1(\widetilde{\Psi},1) \nnb \\
%\ar {i\over 48 \sqrt{3} M^4} (m_d \dd + m_s \sp) (1-\beta) m_V^3
%f_V^\parallel \widetilde{i}_4({\Bbb A}) \nnb \\
%\ek {i\over 108 \sqrt{3} M^4} (m_d \dd + m_s \sp) (1-\beta) m_0^2 m_V
%f_V^\parallel \widetilde{i}_4(\Bbb{B}) \nnb \\
%\ek {i\over 432 \sqrt{3} \pi^2 M^4} [6 m_d - 3 m_s (3+\beta)] \GG f_V^\perp
%m_V^2 \widetilde{\widetilde{i}}_4(\Bbb{B}_T) \nnb \\
%\ar {i\over 54 \sqrt{3} M^4} [ \dd (5-\beta) - 2 \sp (5 + 2 \beta)] m_0^2  
%f_V^\perp m_V^2 \widetilde{\widetilde{i}}_4(\Bbb{B}_T) \nnb \\
%\ar {i\over 108\sqrt{3} M^4} (m_d \dd + m_s \sp) (1-\beta) m_0^2 m_V
%f_V^\parallel \widetilde{i}_4(\phi_V^\parallel) \nnb \\
%\ek {i\over 216\sqrt{3} M^4} (m_d \dd + m_s \sp) (1-\beta) m_0^2 m_V    
%f_V^\parallel \psi_{3;V}^\perp(u_0) \nnb \\
%
%
\ar {f_V^\parallel m_V \over 72\sqrt{3} M^2} (1-\beta) \Big\{
4 [m_d \dd i_2({\cal A},v) - m_s \sp i_2({\cal A},1)] \nnb \\
\ek 4 [m_d \dd - 2 m_s \sp] i_2({\cal V},v) - 4 m_s \sp i_2({\cal V},1) \nnb \\
\ek 3 [m_d \dd + m_s \sp]  [2 \widetilde{i}_4(\Bbb{B})-
2 \widetilde{i}_4(\phi_V^\parallel) + \psi_{3;V}^\perp(u_0)] \Big\} \nnb \\
\ek {f_V^\perp m_V^2 \over 9\sqrt{3} M^2} \Big\{
(1+\beta) [\dd - 2 \sp] [i_1({\cal T},1) - i_1({\cal T}_4,1)] \nnb \\
\ek [2 \dd - \sp (3+\beta)] [i_1({\cal T},v) - i_1({\cal T}_4,v) +
3 \widetilde{\widetilde{i}}_4(\Bbb{B}_T)] \Big\}\nnb \\
%
%BU KISIM YUKARIYA YAZILDI(11)
%\ek {i\over 18\sqrt{3} M^2} [m_d \dd i_2({\cal A},v) - m_s \sp i_2({\cal A},1)]
%(1-\beta) m_V f_V^\parallel \nnb \\
%\ar {i\over 18\sqrt{3} M^2} [(m_d \dd - 2 m_s \sp) i_2({\cal V},v) +
%m_s \sp i_2({\cal V},1)] (1-\beta) m_V f_V^\parallel \nnb \\
%\ar {i\over 9\sqrt{3} M^2} \{(\dd - 2 \sp) (1+\beta) i_1({\cal T},1) -
%[2 \dd - \sp (3+\beta)] i_1({\cal T},v)\} m_V^2 f_V^\perp \nnb \\
%\ek {2 i\over 9\sqrt{3} M^2} \{(\dd - 2 \sp) (1+\beta) i_1({\cal T}_4,1) -
%[2 \dd - \sp (3+\beta)] i_1({\cal T}_4,v)\} m_V^2 f_V^\perp \nnb \\
%\ek {i\over 3 \sqrt{3} M^2} [2 \dd - \sp (3+\beta)]
%m_V^2 f_V^\perp \widetilde{\widetilde{i}}_4(\Bbb{B}_T) \nnb \\
%\ar {i\over 12 \sqrt{3} M^2} (m_d \dd + m_s \sp) (1-\beta)
%m_V f_V^\parallel [\widetilde{i}_4(\Bbb{B})-
%\widetilde{i}_4(\phi_V^\parallel)] \nnb \\
%\ar {i\over 24 \sqrt{3} M^2} (m_d \dd + m_s \sp) (1-\beta)
%m_V f_V^\parallel \psi_{3;V}^\perp(u_0) \nnb \\
%
%
\ek {f_V^\parallel M_V M^2 \over 96 \sqrt{3} \pi^2} (1-\beta)
[4 i_2({\cal A},1-v) +
4 i_2({\cal V},1-v) + 2 \widetilde{i}_4(\Bbb{B}) -
2 \widetilde{i}_4(\phi_V^\parallel) + 
\psi_{3;V}^\perp(u_0) ]\nnb \\
%
%BU KISIM YUKARIYA YAZILDI(12)
%\ar {i\over 24 \sqrt{3} \pi^2} M^2 (1-\beta)
%m_V f_V^\parallel i_2({\cal A},1-v) \nnb \\
%\ar {i\over 24 \sqrt{3} \pi^2} M^2 (1-\beta)
%m_V f_V^\parallel i_2({\cal V},1-v) \nnb \\
%\ar {i\over 48 \sqrt{3} \pi^2} M^2 (1-\beta)
%m_V f_V^\parallel \widetilde{i}_4(\Bbb{B}) \nnb \\
%\ek {i\over 48 \sqrt{3} \pi^2} M^2 (1-\beta)
%m_V f_V^\parallel \widetilde{i}_4(\phi_V^\parallel) \nnb \\
%\ar {i\over 96 \sqrt{3} \pi^2} M^2 (1-\beta)
%m_V f_V^\parallel \psi_{3;V}^\perp(u_0) \nnb \\
%
%
\ek {f_V^\perp m_V^2 \over 24 \sqrt{3} \pi^2} \Big\{
2 [m_d (1-3 \beta) + m_s (1+3 \beta)] i_1({\cal T},1) +
4 [2 m_d \beta - m_s (1+3 \beta)] i_1({\cal T}_4,1) \nnb \\
\ar [2 m_d - m_s (3+\beta)] [2 i_1({\cal T},v) -
4 i_1({\cal T}_4,v) + 3 \widetilde{\widetilde{i}}_4(\Bbb{B}_T)] - 
4 m_d (1-\beta) i_1({\cal T}_3,1) \Big\} \nnb \\
\ar {f_V^\parallel m_V^3 \over 192 \sqrt{3} \pi^2} (1-\beta)
\Big\{ 8 [4 i_0(\Psi,1-2 v) + 4 i_0(\widetilde{\Psi},1) +
i_1(\Phi,v) - i_1(\widetilde{\Phi},1-3 v) \nnb \\
\ek i_1(\Psi,1-2 v)] - 3 \widetilde{i}_4({\Bbb A}) \Big\}~, \\ \nnb 
%
%BU KISIM YUKARIYA YAZILDI(13)
%\ar {i\over 12 \sqrt{3} \pi^2} \{ [m_d (1-3 \beta) + m_s (1+3 \beta)]
%i_1({\cal T},1) + [2 m_d - m_s (3+\beta)] i_1({\cal T},v) \} 
%m_V^2 f_V^\perp \nnb \\
%\ek {i\over 6 \sqrt{3} \pi^2} m_d (1-\beta) m_V^2 f_V^\perp 
%i_1({\cal T}_3,1) \nnb \\
%\ar {i\over 6 \sqrt{3} \pi^2} \{ [2 m_d \beta - m_s (1+3 \beta)] i_1({\cal T}_4,1)
%- [2 m_d - m_s (3+\beta)] i_1({\cal T}_4,v) \} m_V^2 f_V^\perp \nnb \\
%\ek {i\over 6 \sqrt{3} \pi^2} (1-\beta) m_V^3 f_V^\parallel
%[i_0(\Psi,1-2 v) + i_0(\widetilde{\Psi},1)] \nnb \\
%\ek {i\over 24 \sqrt{3} \pi^2} (1-\beta) m_V^3 f_V^\parallel
%i_1(\Phi,v) \nnb \\
%\ar {i\over 24 \sqrt{3} \pi^2} (1-\beta) m_V^3 f_V^\parallel
%i_1(\widetilde{\Phi},1-3 v) \nnb \\
%\ar {i\over 24 \sqrt{3} \pi^2} (1-\beta) m_V^3 f_V^\parallel
%i_1(\Psi,1-2 v) \nnb \\
%\ar {i\over 64 \sqrt{3} \pi^2} (1-\beta) m_V^3 f_V^\parallel
%\widetilde{i}_4({\Bbb A}) \nnb \\
%\ar {i\over 8 \sqrt{3} \pi^2} [2 m_d - m_s (3+\beta)] m_V^2 f_V^\perp
%\widetilde{\widetilde{i}}_4(\Bbb{B}_T) \\ \nnb \\ \nnb 
\eaeeq

\baeeq
\label{e102}
\Pi_1^{(3)} \es
- {M^4 \over 96 \sqrt{3} \pi^2} \Big\{f_V^\parallel m_V (1-\beta)
[2 i_3({\cal A},1-v) - 2  i_3({\cal V},1-v) - 2 \widetilde{i}_4(\Bbb{B}) +
2 \widetilde{i}_4(\phi_V^\parallel) \nnb \\
\ar 5 \psi_{3;V}^\perp(u_0)] -
6 f_V^\perp [2 \beta m_d +(1+\beta) m_s] \phi_V^\perp(u_0) \Big\} \nnb \\
%
%BU KISIM YUKARIYA YAZILDI(14)
%{i\over 48 \sqrt{3} \pi^2} M^4 (1-\beta) f_V^\parallel m_V
%i_3({\cal A},1-v) \nnb \\
%\ek {i\over 48 \sqrt{3} \pi^2} M^4 (1-\beta) f_V^\parallel m_V
%i_3({\cal V},1-v) \nnb \\
%\ek {i\over 48 \sqrt{3} \pi^2} M^4 (1-\beta) f_V^\parallel m_V
%\widetilde{i}_4(\Bbb{B}) \nnb \\
%\ar {i\over 48 \sqrt{3} \pi^2} M^4 (1-\beta) f_V^\parallel m_V
%\widetilde{i}_4(\phi_V^\parallel) \nnb \\
%\ek {i\over 16 \sqrt{3} \pi^2} M^4 f_V^\perp [2 \beta m_d +
%(1+\beta) m_s] \phi_V^\perp(u_0) \nnb \\
%\ar {5 i\over 96 \sqrt{3} \pi^2} M^4 (1-\beta) f_V^\parallel m_V
%\psi_{3;V}^\perp(u_0) \nnb \\
%
\ar \Bigg(\gamma_E - \ln {M^2\over \Lambda^2}\Bigg) \Bigg\{
- {f_V^\perp m_V^2 M^2 \over 24 \sqrt{3} \pi^2} \Big(
[m_s (1+3\beta)-2 m_d \beta] i_2({\cal S},1) \nnb \\
\ek [m_s (3+\beta)-2 m_d] [ i_2(\widetilde{\cal S},1) - i_2({\cal T}_4,1) ]
+ 2 [m_d (1-\beta) - m_s (2+\beta)]i_2({\cal T}_1,1) \nnb \\
\ar 2 [m_d (1-\beta) + m_s (1+2 \beta)] i_2({\cal T}_2,1) -
[2 m_d + m_s (1+\beta)] i_2({\cal T}_3,1) \Big)\nnb \\
\ar {f_V^\perp \GG \over 192 \sqrt{3} \pi^2 M^2} \Big( 2 m_V^2 
[2 m_d - m_s (3+\beta)] \widetilde{\widetilde{i}}_4(\Bbb{B}_T) +
M^2 [2 m_d \beta + m_s (1+\beta)] \phi_V^\perp(u_0) \Big) \Bigg\} \nnb \\ 
%
%BU KISIM YUKARIYA YAZILDI(15)
%\ar \Bigg(\gamma_E - \ln {M^2\over \Lambda^2}\Bigg) \Bigg\{
%{i\over 24 \sqrt{3} \pi^2} M^2 [m_s (1+3\beta)-2 m_d \beta] f_V^\perp
%m_V^2 i_2({\cal S},1) \nnb \\
%\ek {i\over 24 \sqrt{3} \pi^2} M^2 [m_s (3+\beta)-2 m_d] f_V^\perp
%m_V^2 i_2(\widetilde{\cal S},1) \nnb \\
%\ar {i\over 12 \sqrt{3} \pi^2} M^2 [m_d (1-\beta) - m_s (2+\beta)]
%f_V^\perp m_V^2 i_2({\cal T}_1,1) \nnb \\
%\ar {i\over 12 \sqrt{3} \pi^2} M^2 [m_d (1-\beta) + m_s (1+2 \beta)]
%f_V^\perp m_V^2 i_2({\cal T}_2,1) \nnb \\
%\ek {i\over 24 \sqrt{3} \pi^2} M^2 [2 m_d + m_s (1+\beta)]
%f_V^\perp m_V^2 i_2({\cal T}_3,1) \nnb \\
%\ek {i\over 24 \sqrt{3} \pi^2} M^2 [2 m_d - m_s (3+\beta)]  
%f_V^\perp m_V^2 i_2({\cal T}_4,1) \nnb \\
%
%\ek {i\over 96 \sqrt{3} \pi^2 M^2} [2 m_d - m_s (3+\beta)]  
%\GG f_V^\perp m_V^2 \widetilde{\widetilde{i}}_4(\Bbb{B}_T) \nnb \\
%\ek {i\over 192 \sqrt{3} \pi^2} \GG [2 m_d \beta + m_s (1+\beta)]
%f_V^\perp \phi_V^\perp(u_0)
%\Bigg\} \nnb \\
%
\ek {f_V^\parallel m_V \GG \over 144\sqrt{3} M^6} 
\beta [m_d \sp + m_s \dd] (m_0^2 + 2 M^2 )\psi_{3;V}^\perp(u_0) \nnb \\
%
%BU KISIM YUKARIYA YAZILDI(16)
%\ar {i\over 144\sqrt{3} M^6} m_0^2 m_V \GG f_V^\parallel
%(m_d \sp + m_s \dd) \beta \psi_{3;V}^\perp(u_0) \nnb \\
%\ar {i\over 72 \sqrt{3} M^4} \GG \beta (m_d \sp + m_s \dd)    
%f_V^\parallel m_V \psi_{3;V}^\perp(u_0) \nnb \\
%
\ar {f_V^\perp m_V^2 \GG \over 2304 \sqrt{3} \pi^2 M^2} \Big\{
[2 m_d \beta + m_s (1+\beta)] [3\Bbb{A}_T(u_0) - 
4 i_2({\cal T}_3,1-2 v)] \nnb \\
\ek 8 [m_d (1+\beta) - m_s] [i_2(\widetilde{\cal S},1- 2 v) - 
i_2({\cal T}_4,1-2 v)] - 8 (2 m_d - m_s)i_2({\cal S},1) \nnb \\
\ar 4 [m_s (3+\beta) - 2 m_d] i_2({\cal T}_1,1) +
16 [m_d (1+\beta) - m_s] i_2({\cal T}_1,v)  \nnb \\
\ek 4 [2 m_d \beta - m_s (1+3 \beta)] i_2({\cal T}_2,1) - 
8 [2 \beta m_d + m_s (1+\beta)] i_2({\cal T}_2,v) \Big\} \nnb \\
%
%BU KISIM YUKARIYA YAZILDI(17)
%\ek {i\over 768 \sqrt{3} \pi^2 M^2} \GG [2 m_d \beta + m_s (1+\beta)]
%f_V^\perp m_V^2 \Bbb{A}_T(u_0) \nnb \\
%\ar {i \over 288 \sqrt{3} \pi^2 M^2} \GG (2 m_d - m_s) \beta f_V^\perp
%m_V^2 i_2({\cal S},1) \nnb \\
%\ar {i\over 288 \sqrt{3} \pi^2 M^2} \GG [m_d (1+\beta) - m_s]
%i_2(\widetilde{\cal S},1- 2 v) f_V^\perp m_V^2 \nnb \\
%\ek {i\over 576 \sqrt{3} \pi^2 M^2} \GG \{ 
%[m_s (3+\beta) - 2 m_d] i_2({\cal T}_1,1) \nnb \\
%\ar 4 [m_d (1+\beta) - m_s] i_2({\cal T}_1,v) \} f_V^\perp m_V^2 \nnb \\
%\ar {i\over 576 \sqrt{3} \pi^2 M^2} \GG f_V^\perp m_V^2 
%\{[2 m_d \beta - m_s (1+3 \beta)] i_2({\cal T}_2,1) \nnb \\
%\ar 2 [2 \beta m_d + m_s (1+\beta)] i_2({\cal T}_2,v) \} \nnb \\
%\ar {i\over 576 \sqrt{3} \pi^2 M^2} \GG f_V^\perp m_V^2
%[2 m_d \beta + m_s (1+\beta)] i_2({\cal T}_3,1-2 v) \nnb \\
%\ek {i\over 288 \sqrt{3} \pi^2 M^2} \GG [m_d (1+\beta) - m_s] f_V^\perp
%m_V^2 i_2({\cal T}_4,1-2 v) \nnb \\
%
\ek {f_V^\parallel m_V^3 \over 288 \sqrt{3} M^2} (1-\beta)
\Big\{[m_d \dd - m_s \sp][8 i_1(\Phi,1) + 9 i_1(\widetilde{\Psi},1)] \nnb \\
\ek 2 [m_d \dd + m_s \sp] [4 i_1(\widetilde{\Phi},1) + 4 i_1(\Psi,1-2 v) +
3 \widetilde{i}_4({\Bbb A})] + 16 [2 m_d \dd - m_s \sp] i_1(\Phi,v) \nnb \\
\ar 16 [m_d \dd + 2 m_s \sp] i_1(\widetilde{\Phi},v) +
16 [m_d \dd i_2({\cal A},1) - m_s \sp i_2({\cal A},v)] \nnb \\
\ek 16 [m_d \dd i_2({\cal V},1) - m_s \sp i_2({\cal V},v)] \Big\}\nnb \\
\ek {f_V^\parallel m_V m_0^2 \over 216 \sqrt{3} M^2} \Big\{
2 (1-\beta) [m_d \dd + m_s \sp] [\widetilde{i}_4(\Bbb{B}) -
\widetilde{i}_4(\phi_V^\parallel)] \nnb \\
\ek \Big(\sp [ 2 m_d (1+\beta) -
m_s (1-\beta)] - \dd [ m_d (1-\beta) - 2 m_s (1+\beta)] \Big)
\psi_{3;V}^\perp(u_0) \Big\} \nnb \\
\ek {f_V^\perp m_V^2 \over 432 \sqrt{3} \pi^2 M^2} \Big\{
\GG [6 m_d - 3 m_s (3+\beta)]
- 8 \pi^2 m_0^2 [ \dd (5-\beta) - 2 \sp (5 + 2 \beta)] \Big\} 
\widetilde{\widetilde{i}}_4(\Bbb{B}_T) \nnb \\
\ek {f_V^\perp m_V^2 M^2 \over 96 \sqrt{3} \pi^2} \Big\{
[2 m_d \beta + m_s (1+\beta)] [3 \Bbb{A}_T(u_0) - 4 i_2(\widetilde{\cal
S},1) - 4 i_2({\cal T},1) \nnb \\
\ar 8 i_2({\cal T}_3,v) + 4 i_2({\cal T}_4,1) ]
+ 16 [m_d (1+\beta) - m_s] [i_2(\widetilde{\cal S},v) + i_2({\cal T}_1,v) 
- i_2({\cal T}_4,v)] \nnb \\
\ek 4 [6 m_d \beta - m_s (1+5 \beta)] i_2({\cal S},1) -
8 (m_d+m_s) (1+\beta) i_2({\cal T}_3,1)  \Big\} \nnb \\
%
%BU KISIM YUKARIYA YAZILDI(19)
%\ar {i\over 32 \sqrt{3} \pi^2} M^2 [2 m_d \beta + m_s (1+\beta)] f_V^\perp
%m_V^2 \Bbb{A}_T(u_0) \nnb \\
%\ek {i\over 24 \sqrt{3} \pi^2} M^2 [6 m_d \beta - m_s (1+5 \beta)]
%f_V^\perp m_V^2 i_2({\cal S},1) \nnb \\
%\ek {i\over 24 \sqrt{3} \pi^2} M^2 [2 m_d \beta + m_s (1+\beta)]
%f_V^\perp m_V^2 i_2(\widetilde{\cal S},1) \nnb \\
%\ar {i\over 6 \sqrt{3} \pi^2} M^2 [m_d (1+\beta) - m_s]
%f_V^\perp m_V^2 i_2(\widetilde{\cal S},v) \nnb \\
%\ek {i\over 24 \sqrt{3} \pi^2} M^2 [2 m_d \beta + m_s (1+\beta)]
%f_V^\perp m_V^2 i_2({\cal T},1) \nnb \\
%\ar  {i\over 6 \sqrt{3} \pi^2} M^2 [m_d (1+\beta) - m_s]
%f_V^\perp m_V^2 i_2({\cal T}_1,v) \nnb \\
%\ek {i\over 12 \sqrt{3} \pi^2} M^2 \{ (m_d+m_s) (1+\beta) i_2({\cal T}_3,1) -
%[2 m_d \beta + m_s (1+\beta)] i_2({\cal T}_3,v) \}
%m_V^2 f_V^\perp \nnb \\
%\ar {i\over 24 \sqrt{3} \pi^2} M^2 \{ [2 m_d \beta + m_s (1+\beta)] i_2({\cal T}_4,1)
%- 4 [m_d (1+\beta) - m_s] i_2({\cal T}_4,v) \}
%m_V^2 f_V^\perp \nnb \\
%
\ek {f_V^\parallel m_V^3 M^2 \over 192 \sqrt{3} \pi^2} (1-\beta)
\Big\{8 [i_1(\Phi,v) - i_1(\widetilde{\Phi},1-3 v) - i_1(\Psi,1-2 v) \nnb \\
\ek i_2({\cal V},1-v) + i_2({\cal A},1-v)] - 3 \widetilde{i}_4({\Bbb A})
\Big\} \nnb \\
\ar {f_V^\perp M^2 \over 24 \sqrt{3} \pi^2} \Big\{
3 m_V^2 [2 m_d - m_s (3+\beta)] \widetilde{\widetilde{i}}_4(\Bbb{B}_T) -
4 \pi^2 [2 \dd \beta + \sp (1+\beta)] \phi_V^\perp(u_0) \Big\} \nnb \\ 
%
%BU KISIM YUKARIYA YAZILDI(20)
%\ar {i\over 24 \sqrt{3} \pi^2} M^2 (1-\beta) f_V^\parallel m_V^3
%i_1(\Phi,v) \nnb \\
%\ek {i\over 24 \sqrt{3} \pi^2} M^2 (1-\beta) f_V^\parallel m_V^3
%i_1(\widetilde{\Phi},1-3 v) \nnb \\
%\ek {i\over 24 \sqrt{3} \pi^2} M^2 (1-\beta) f_V^\parallel m_V^3
%i_1(\Psi,1-2 v) \nnb \\
%\ek {i\over 24 \sqrt{3} \pi^2} M^2 (1-\beta) f_V^\parallel m_V^3
%i_2({\cal V},1-v) \nnb \\
%\ar {i\over 24 \sqrt{3} \pi^2} M^2 (1-\beta) f_V^\parallel m_V^3
%i_2({\cal A},1-v) \nnb \\
%\ek {i\over 64 \sqrt{3} \pi^2} M^2 (1-\beta) f_V^\parallel m_V^3
%\widetilde{i}_4({\Bbb A}) \nnb \\
%\ek {i\over 8 \sqrt{3} \pi^2} M^2 [2 m_d - m_s (3+\beta)] f_V^\perp
%m_V^2 \widetilde{\widetilde{i}}_4(\Bbb{B}_T) \nnb \\
%\ar {i\over 6 \sqrt{3}} M^2 [2 \dd \beta + \sp (1+\beta)] f_V^\perp
%\phi_V^\perp(u_0) \nnb \\
%
\ek {f_V^\parallel m_V \over 36 \sqrt{3}} (1-\beta) \Big\{
[m_d \dd i_3({\cal A},1) - m_s \sp i_3({\cal A},v)] -
[m_d \dd i_3({\cal V},1) - m_s \sp i_3({\cal V},v)] \nnb \\
\ek 3 [m_d \dd + m_s \sp] [\widetilde{i}_4(\Bbb{B}) - 
\widetilde{i}_4(\phi_V^\parallel)] \Big\} \nnb \\
\ek  {f_V^\parallel m_V \over 24 \sqrt{3} \pi^2} \Big\{
[m_d (1-\beta) + 8 m_s \beta] \dd + [8 \beta m_d + m_s (1-\beta)] \sp
\Big\} \psi_{3;V}^\perp(u_0) \nnb \\
\ek {f_V^\perp m_V^2 \over 72 \sqrt{3}} \Big\{
[2 \beta \dd + (1+\beta) \sp] [4 i_2({\cal T}_3,1-2 v) - 3 \Bbb{A}_T(u_0)] 
+ 24 [2 \dd - \sp (3+\beta)] \widetilde{\widetilde{i}}_4(\Bbb{B}_T) \nnb \\
\ar 8 [\dd (1+\beta) - \sp] [ i_2(\widetilde{\cal S},1-2 v) -
i_2({\cal T}_4,1-2 v)] + 8 \beta [2 \dd - \sp] i_2({\cal S},1)
\Big\} \nnb \\
\ek {f_V^\perp m_V^2 \over 18\sqrt{3}} \Big\{
[2 \dd - (3+\beta) \sp] i_2({\cal T}_1,1) - 4 [\dd (1+\beta) -\sp]
i_2({\cal T}_1,v) \nnb \\
\ar [2 \beta \dd - (1+3 \beta) \sp] i_2({\cal T}_2,1) + 2 [2 \dd \beta +
(1+\beta) \sp] i_2({\cal T}_2,v) \Big\} \nnb \\
\ek {f_V^\perp \over 864 \sqrt{3} \pi^2} \Big\{
3 \GG [2 m_d \beta + m_s (1+\beta)] -
40 \pi^2 m_0^2 [2 \dd \beta + \sp (1+\beta)] \Big\} \phi_V^\perp(u_0)~.  
\eaeeq

The functions $i_n$, $\widetilde{i}_4$ and
$\widetilde{\widetilde{i}}_4$ are defined as
\baeeq
\label{nolabel}
i_0(\phi,f(v)) \es \int {\cal D}\alpha_i \int_0^1 dv
\phi(\alpha_{\bar{q}},\alpha_q,\alpha_g) f(v) (k-u_0) \theta(k-u_0)~, \nnb \\
i_1(\phi,f(v)) \es \int {\cal D}\alpha_i \int_0^1 dv
\phi(\alpha_{\bar{q}},\alpha_q,\alpha_g) f(v) \theta(k-u_0)~, \nnb \\
i_2(\phi,f(v)) \es \int {\cal D}\alpha_i \int_0^1 dv
\phi(\alpha_{\bar{q}},\alpha_q,\alpha_g) f(v) \delta(k-u_0)~, \nnb \\
i_3(\phi,f(v)) \es \int {\cal D}\alpha_i \int_0^1 dv
\phi(\alpha_{\bar{q}},\alpha_q,\alpha_g) f(v) \delta^\prime(k-u_0)~, \nnb \\
\widetilde{i}_4(f(u)) \es \int_{u_0}^1 du f(u)~, \nnb \\
\widetilde{\widetilde{i}}_4(f(u)) \es \int_{u_0}^1 du (u-u_0) f(u)~, \nnb
\eaeeq
where 
\baeeq
k = \alpha_q + \alpha_g \bar{v}~,~~~~~u_0={M_1^2 \over M_1^2
+M_2^2}~,~~~~~M^2={M_1^2 M_2^2 \over M_1^2
+M_2^2}~.\nnb
\eaeeq

Note that $\Pi^{(i)}(u,d,s),~(i=u,d,s)$ do not depend explicitly on the $u$
quark properties $m_u$ and $\uu$. The dependence is implicit in the meson
distribution amplitudes and leptonic constants.

\eAPP

\newpage

\begin{figure}
\vskip 3. cm
    \includegraphics{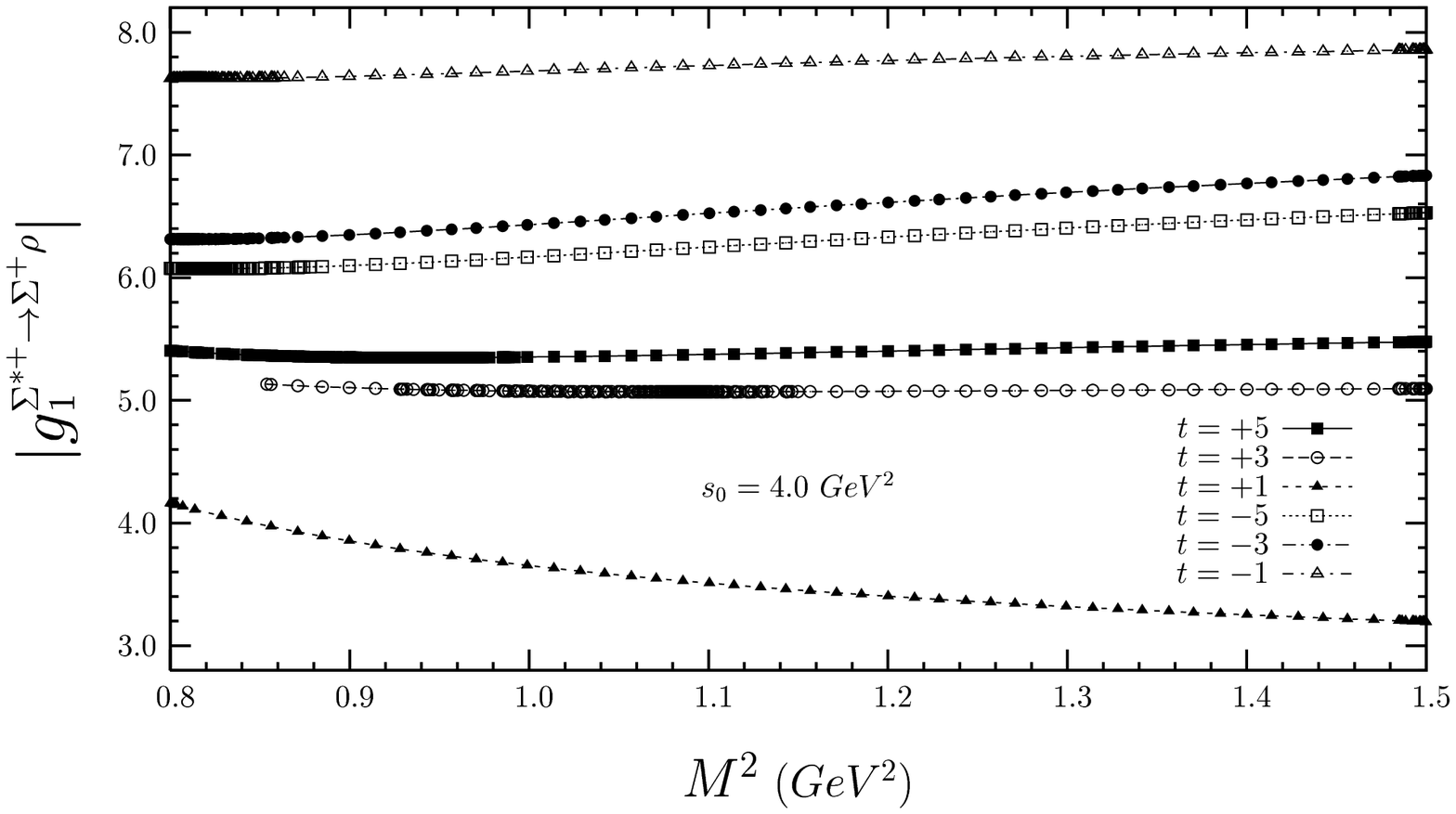}
\vskip 6.3cm
\caption{}
%\begin{center}
%{\bf Fig. 1--a}
%\end{center}
\end{figure}

\begin{figure}
\vskip 4.0 cm
    \includegraphics{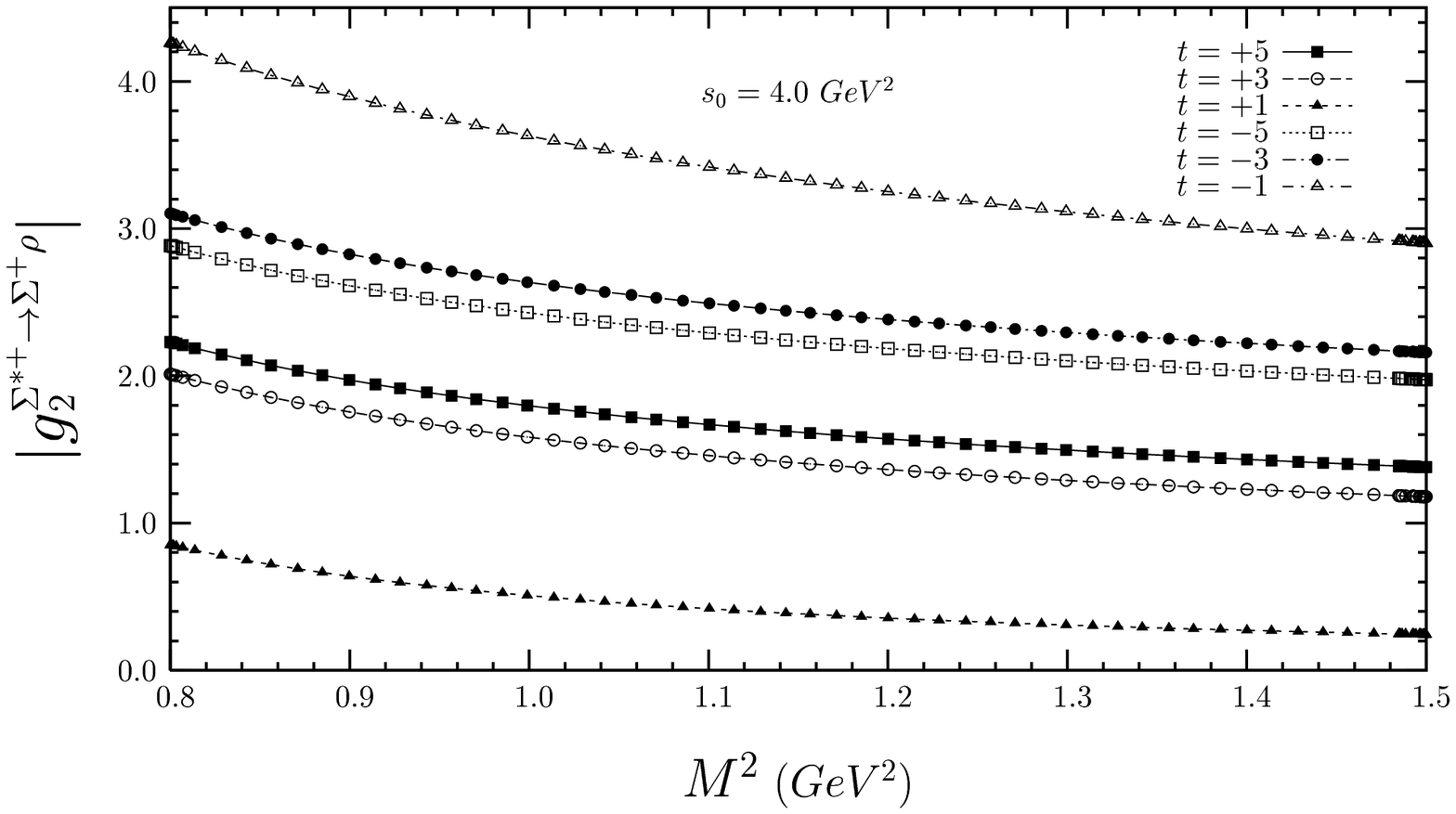}
\vskip 6.3 cm
\caption{}
%\begin{center}
%{\bf Fig. 1--b}
%\end{center}
\end{figure}

\newpage

\begin{figure}
\vskip 3. cm
    \includegraphics{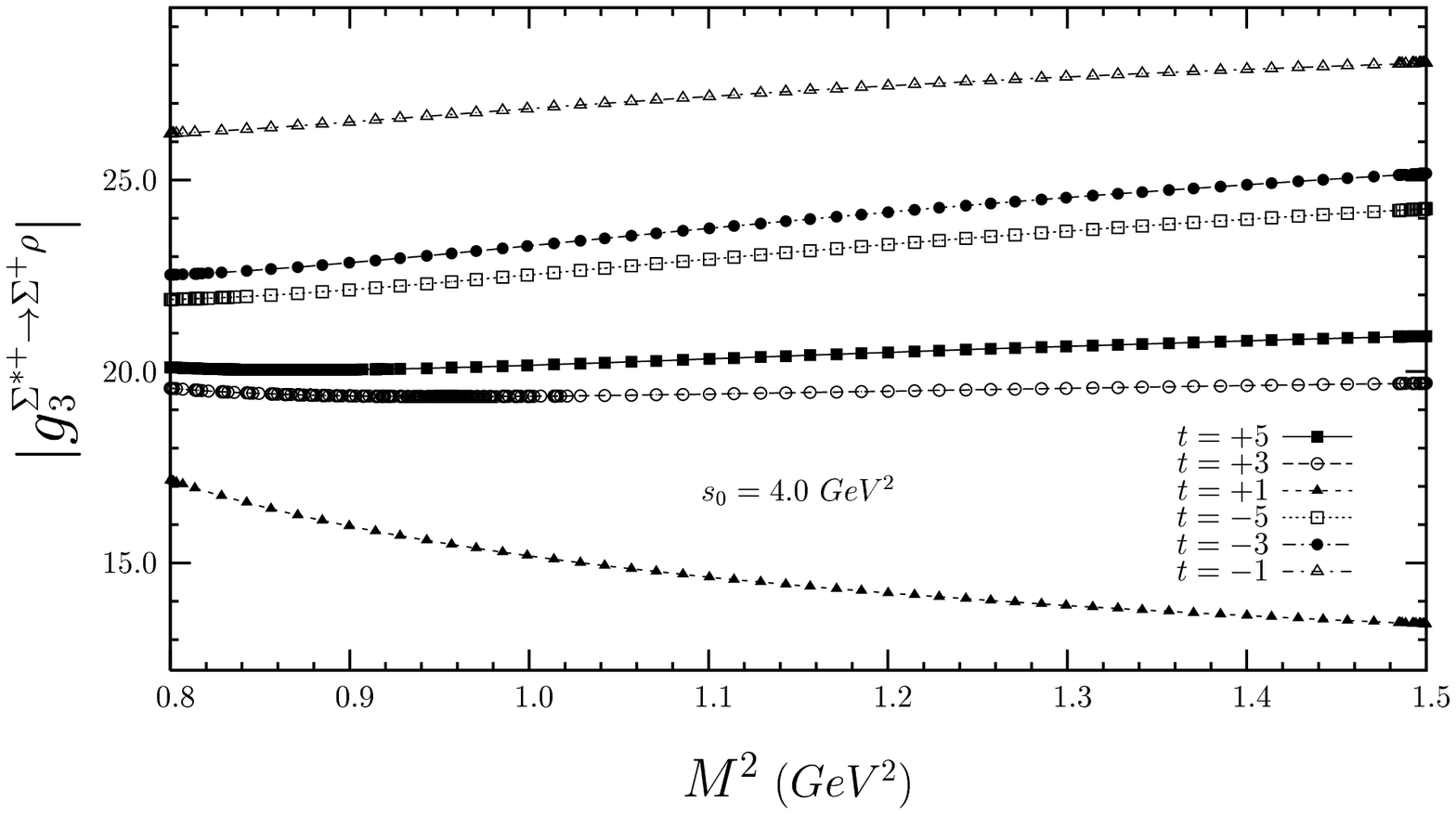}
\vskip 6.3cm
\caption{}
%\begin{center}
%{\bf Fig. 1--a}
%\end{center}
\end{figure}

\begin{figure}
\vskip 4.0 cm
    \includegraphics{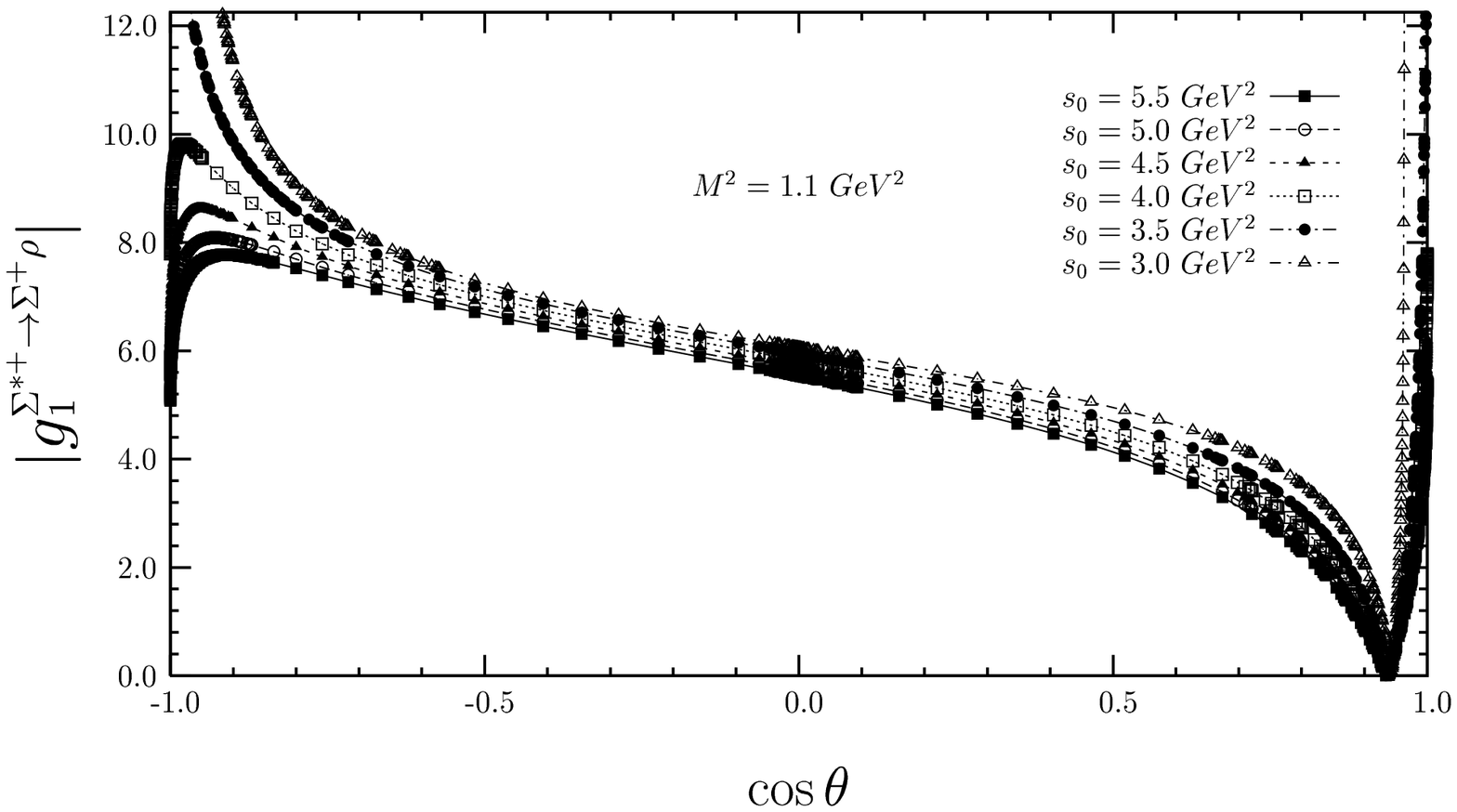}
\vskip 6.3 cm
\caption{}
%\begin{center}
%{\bf Fig. 1--b}
%\end{center}
\end{figure}

\newpage

\begin{figure}
\vskip 3. cm
    \includegraphics{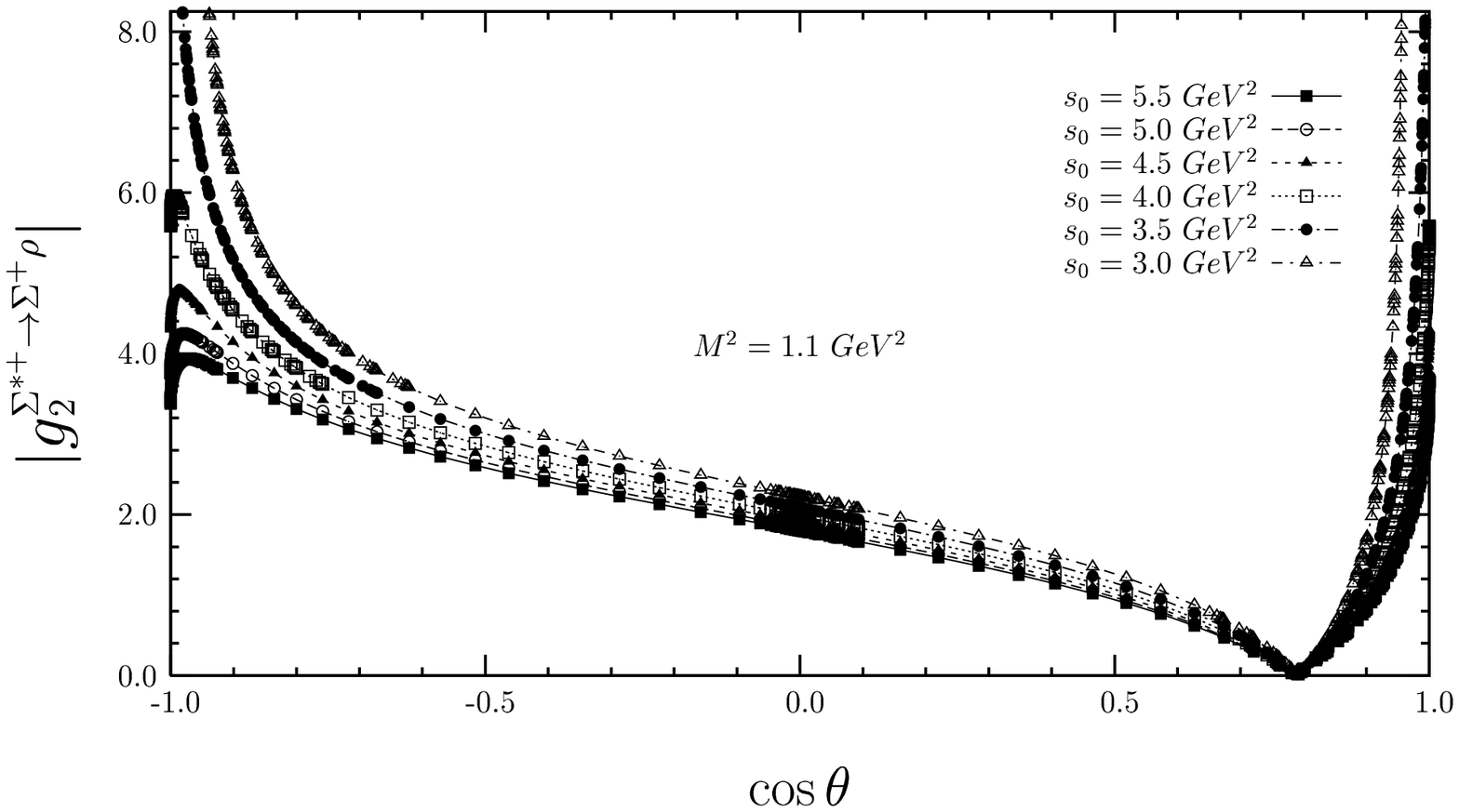}
\vskip 6.3cm
\caption{}
%\begin{center}
%{\bf Fig. 1--a}
%\end{center}
\end{figure}

\begin{figure}
\vskip 4.0 cm
    \includegraphics{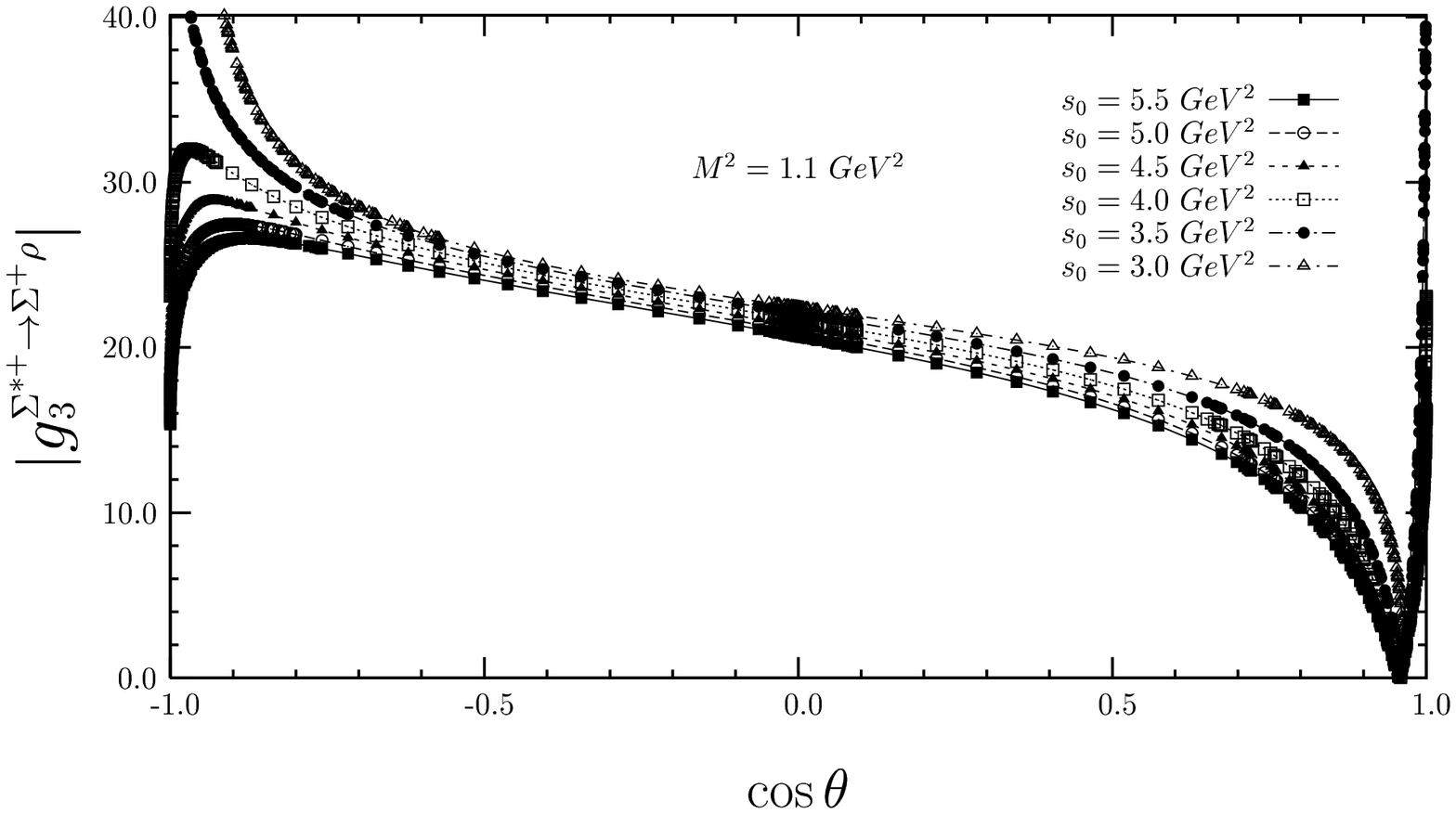}
\vskip 6.3 cm
\caption{}
%\begin{center}
%{\bf Fig. 1--b}
%\end{center}
\end{figure}

\end{document}